\documentclass{aa}  

\usepackage{graphicx}

\usepackage{amsmath}

\usepackage{txfonts}
\usepackage{natbib}
\usepackage{booktabs}
\usepackage{titlesec}
\usepackage{bm}
\usepackage{booktabs}
\usepackage{hyperref}
\hypersetup{
    colorlinks = true,
    allcolors = blue,
}
\titleformat{\paragraph}
{\normalfont\normalsize\bfseries}{\theparagraph}{1em}{}
\titlespacing*{\paragraph}
{0pt}{3.25ex plus 1ex minus .2ex}{1.5ex plus .2ex}

 \bibpunct{(}{)}{;}{a}{}{,} 
\begin{document}

   \title{Catalogue of new Herbig Ae/Be and classical Be stars
   \thanks{Tables \ref{Table2}, \ref{Table3}, Sample of Study with probabilities (Sect. \ref{S_out}), and table of sources that belong to either category (Sect. \ref{Sect_to_either}) are only available in electronic form at the CDS via anonymous ftp to cdsarc.u-strasbg.fr (130.79.128.5), via http://cdsweb.u-strasbg.fr/cgi-bin/qcat?J/A+A/, or via the \href{https://starry-project.eu/publications/}{STARRY webpage}.}}

   \subtitle{A machine learning approach to \textit{Gaia} DR2}

   \author{M. Vioque
          \inst{1,2}
          \and
          R.D. Oudmaijer\inst{1}
          \and
          M. Schreiner\inst{3}
          \and          
          I. Mendigutía\inst{4}
          \and          
          D. Baines\inst{5}
          \and          
          N. Mowlavi\inst{6}      
          \and
          R. P\'erez-Mart\'inez\inst{2}          
          }

   \institute{School of Physics and Astronomy, University of Leeds, Leeds LS2 9JT, UK\\
   \email{\href{mailto:pymvdl@leeds.ac.uk}{pymvdl@leeds.ac.uk}}
   \and
   Ingenier\'ia de Sistemas para la Defensa de Espa\~na (Isdefe), XMM/Newton Science Operations Centre, ESA-ESAC Campus, PO Box 78, 28691 Villanueva de la Ca\~nada, Madrid, Spain
   \and
   Desupervised, Desupervised ApS, Njalsgade 76, 2300 Copenhagen, Denmark
   \and
   Centro de Astrobiología (CSIC-INTA), Departamento de Astrofísica, ESA-ESAC Campus, PO Box 78, 28691 Villanueva de la Cañada, Madrid, Spain 
   \and   
   Quasar Science Resources for ESA-ESAC, ESAC Science Data Centre,  PO Box 78, 28691 Villanueva de la Ca\~nada, Madrid, Spain   
   \and
   Department of Astronomy, University of Geneva, Ch. des Maillettes 51, 1290 Versoix, Switzerland\\}

   \date{Accepted for publication in Astronomy \& Astrophysics. Received 14 February 2020 / Accepted 27 April 2020}

 
  \abstract
   {The intermediate-mass pre-main sequence Herbig Ae/Be stars are key to understanding the differences in formation mechanisms between low- and high-mass stars. The study of the general properties of these objects is hampered by the lack of a well-defined, homogeneous sample, and because few and mostly serendipitously discovered sources are known.}
   {Our goal is to identify new Herbig Ae/Be candidates to create a homogeneous and well defined catalogue of these objects.}
   {We have applied machine learning techniques to 4,150,983 sources with data from \textit{Gaia} DR2, 2MASS, WISE, and IPHAS or VPHAS+. Several observables were chosen to identify new Herbig Ae/Be candidates based on our current knowledge of this class, which is characterised by infrared excesses, photometric variabilities, and H$\alpha$ emission lines. Classical techniques are not efficient for identifying new Herbig Ae/Be stars mainly because of their similarity with classical Be stars, with which they share many characteristics. By focusing on disentangling these two types of objects, our algorithm has also identified new classical Be stars.}
   {We have obtained a large catalogue of 8470 new pre-main sequence candidates and another catalogue of 693 new classical Be candidates with a completeness of $78.8\pm1.4\%$ and $85.5\pm1.2\%$, respectively. Of the catalogue of pre-main sequence candidates, at least 1361 sources are potentially new Herbig Ae/Be candidates according to their position in the Hertzsprung-Russell diagram. In this study we present the methodology used, evaluate the quality of the catalogues, and perform an analysis of their flaws and biases. For this assessment, we make use of observables that have not been accounted for by the algorithm and hence are selection-independent, such as coordinates and parallax based distances. The catalogue of new Herbig Ae/Be stars that we present here increases the number of known objects of the class by an order of magnitude.}
   {}


   \maketitle
%

\section{Introduction}\label{S_Introduction}

Herbig Ae/Be stars (HAeBes) are pre-main sequence (PMS) sources of intermediate-mass (canonically defined as $2M_{\odot}\lesssim M\lesssim10M_{\odot}$, spectral type B, A, and F) that cover the gap between the lower-mass T-Tauri stars and the deeply embedded infrared-bright Massive Young Stellar Objects. HAeBes are thus key for understanding the properties of high-mass star formation. However, a large caveat in all of the studies dedicated to HAeBes is that 273 of them are known (108 in the master list of \citealp{1994A&AS..104..315T}, see \citealp{2018A&A...620A.128V}). This is a very heterogeneous and biased set. In particular, few objects are known at the high-mass end (Herbig Be stars), with many of them having a doubtful nature as they are easily confused with classical Be stars (CBes, rapidly rotating main sequence B stars with Keplerian gas discs, \citealp{2013A&ARv..21...69R}). This situation contrasts with the thousands of T-Tauri stars known in the literature. As a consequence, many open problems involving high-mass star formation suffer from these biases and the lack of completeness.

For example, it is commonly accepted that T-Tauri stars accrete through magnetically-driven flows arising from the protoplanetary disc, which is truncated at a distance of a few stellar-radii (see \citealp{2007prpl.conf..479B}; \citealp{2016ARA&A..54..135H}). However, higher-mass PMS objects have radiative envelopes and hence normally present negligible magnetic fields (\citealp{2013MNRAS.429.1001A}; \citealp{2019A&A...622A..72V}). Therefore, the magnetospheric accretion model probably cannot apply to them. The transition from magnetospheric accretion to the still unknown accretion mechanism for higher-mass PMS objects takes place within the mass range of the Herbig Ae/Be stars. Indeed, near-IR interferometric (e.g. \citealp{2005ApJ...624..832M}), optical- and near-UV spectro-polarimetric (e.g. \citealp{2017MNRAS.472..854A}), and spectro-photometric observations (e.g. \citealp{2011A&A...535A..99M}; \citealp{2015MNRAS.453..976F}; \citealp{2020MNRAS.493..234W}) have shown that the lower mass Herbig Ae stars show accretion signatures consistent with T-Tauri stars, whereas Herbig Be stars appear to be inconsistent with magnetospheric accretion. A large caveat in these studies is that they do not include the less evolved sources in high-mass PMS tracks (most Herbig Be stars observed to date are very close to the main sequence, \citealp{2018A&A...620A.128V}), which are obviously of paramount importance for understanding high-mass accretion. In addition, there is observational evidence that points towards differences between the discs of low- and high-mass PMS sources. This can be seen in the amount of infrared excess, which is much lower for high-mass sources (\citealp{2015A&A...576A..52R}; \citealp{2018A&A...620A.128V}; \citealp{2019AJ....157..159A}) or in morphology; for instance, spirals have only been found in early spectral type stars (\citealp{2018A&A...620A..94G}). Similarly, there is a clear observational bias in these results, as so far mostly long-lived, massive discs around low-mass stars have been observed.

Independently, it is known that high-mass stars tend to form in clusters (\citealp{1995AJ....109..280H}; \citealp{1999A&A...342..515T}). Studies of massive field runaway stars have shown that at least a small fraction ($\sim4\%$, \citealp{2005A&A...437..247D}) of O-type stars are formed without a cluster environment. Nonetheless, recent publications question the existence of isolated high-mass star formation (e.g. \citealp{2017ApJ...834...94S}). Again, the scarcity of known high-mass PMS sources makes the statistics non-robust.  

It is thus useful to obtain a large homogeneous and low biased catalogue of new Herbig Ae/Be stars. \textit{Gaia} Data Release 2 (DR2,
\citealp{2016A&A...595A...1G, 2018A&A...616A...1G}) provides a five dimensional astrometric solution for up to $G\lesssim21$ mag (white G band,
described in \citealp{2018A&A...616A...4E}) to over 1.3 billion objects (\citealp{2018A&A...616A...2L}). This large dataset allows for exploitation with statistical learning techniques (as done in e.g.  \citealp{2019MNRAS.487.2522M} or \citealp{2019A&A...626A..80C}; see \citealp{2019arXiv190407248B} for a general description of these techniques into astronomy). In this paper we use an algorithm based on an artificial neural network (ANN) to identify new Herbig Ae/Be stars within \textit{Gaia} DR2. ANNs are supervised learning classifiers, this means that they need to be trained with a list of known sources (training set) that have a set of characteristics (features) and a label (ground truth) that assign them to a certain category (e.g. a stellar class). Once trained, ANNs assign probabilities of belonging to every one of the chosen categories to each input source. The known HAeBes constitute a small, biased, and contaminated set. In order to achieve a good training performance the strategy adopted was to include T-Tauri stars in the training and use an algorithm focusing on the high-mass end. In the resulting catalogue of new PMS candidates, the most massive ones can be further selected by means of the Hertzsprung-Russell (HR) diagram.

The features that feed the ANN need to be relevant for identifying PMS sources. Hence, we want the features to trace the main observational characteristics of PMS sources, which are: infrared (IR) excesses, because of the radiation of the heated up protoplanetary disc, emission lines, that trace the surrounding material close to the forming star, and photometric variability. This PMS variability is caused by the presence of the disc in the line of sight (e.g. dippers, \citealp{1999A&A...349..619B}, or UX Ori type sources, \citealp{2018A&A...620A.128V}), because of episodic accretion events (EX Lup or FU Ori type sources, \citealp{2017ApJ...836...41C}), or pulsations due to internal instability (\citealp{2014Sci...345..550Z}). To feed the algorithm with these characteristics, we use observables belonging to five different surveys: \textit{Gaia} DR2 for variability, 2MASS (\citealp{2006AJ....131.1163S}) and WISE (\citealp{2010AJ....140.1868W}) for near- and mid-IR excess respectively, and IPHAS (\citealp{2005MNRAS.362..753D}; \citealp{2014MNRAS.444.3230B}) and VPHAS+ (\citealp{2014MNRAS.440.2036D}) for H$\alpha$ emission. If HAeBes were unique in these properties, a simple linear separation in the parameter space would suffice for identifying more objects of the class (e.g. in a colour-colour plot). However,  HAeBes share these characteristics with other types of objects, of which classical Be stars stand out, as their outwardly diffusing gaseous discs generate very similar observables (\citealp{2006ApJ...651L..53G}; \citealp{2013A&ARv..21...69R}; \citealp{2017A&A...601A..74K}). Therefore, our ANN-based algorithm spotlights on disentangling these two types of objects, and as a consequence we also find new classical Be candidates. 

The paper is organised as follows: in Sect. \ref{S_parameters}, we describe the observables, features, and the metrics used for evaluating the performance of the algorithm as well as the sources that the algorithm classifies once it is trained. In Sect. \ref{Trainingset} we present the labelled sources used for training the ANN, in Sect. \ref{S_out} we describe and evaluate the output of the algorithm which we analyse in Sect. \ref{Quality assessment}, describing its flaws and biases. Sect. \ref{S_conclusions} summarises the main conclusions. The algorithm itself is detailed in Appx. \ref{S_Description of the pipeline}.

\section{Observables, features, and data}\label{S_parameters}

The features are the individual properties or characteristics that are used by the ANN to learn how to classify new sources. Feature selection is important, as the use of useless features or the lack of very relevant ones for differentiating the categories can heavily affect the performance of the algorithm.

\subsection{Observables}\label{Observables}

As described in Sect. \ref{S_Introduction}, we need observables contained within the catalogues \textit{Gaia} DR2, 2MASS, WISE, and IPHAS and VPHAS+. These catalogues have information in several passbands ranging from the optical to the mid-infrared. We used the following passbands: from \textit{Gaia} DR2, the broad white $G$ band ($0.59\mu m$), and the blue ($G_{BP}$) and red ($G_{RP}$) bands ($0.50\mu m$ and $0.77\mu m$ respectively). A description of the \textit{Gaia} filters can be found in \citet{2018A&A...616A...4E}. From IPHAS and VPHAS+, we used the SLOAN passband $r$ ($0.62\mu m$) 
together with the $H\alpha$ narrow filter ($0.66\mu m$). A description of the IPHAS passbands and associated footprints can be found in \citet{2005MNRAS.362..753D} and \citet{2014MNRAS.444.3230B} (for the second data release that we are using) and in \citet{2014MNRAS.440.2036D} for VPHAS+.  Finally, from 2MASS we used the three passbands $J$, $H$, and $K_{s}$ ($1.24\mu m$, $1.66\mu m$ and $2.16\mu m$ respectively) and from WISE the four passbands $W1$, $W2$, $W3$, and $W4$ ($3.4\mu m$, $4.6\mu m$, $12\mu m$, and $22\mu m$ respectively). These passbands of 2MASS and WISE were obtained from the AllWISE catalogue, which is described in \citet{2013wise.rept....1C}.

It is important when setting up the features to be cautious about introducing unwanted bias regarding the selection we want to perform. An example of an unwanted bias is, for example, to introduce distance as a feature. Most of the known PMS objects are close-by because it is easier to study bright objects. If we introduce a distance dependent feature the algorithm would work with the idea that being close is an intrinsic property of PMS objects, and it would be biased to find PMS objects that are nearby. In addition, if we introduce position dependent features any posterior analysis about the clustering properties of Herbig Ae/Be stars would be biased towards the selected preferred positions of the training data. Therefore, we set the features to be distance and position independent, which implies that most of the observables used are colours. Of course, there are unwanted biases in the resulting catalogues because of the selected features, and they are addressed in Sect. \ref{Quality assessment}. For example, interstellar extinction results in colours that are not strictly distance-independent, and by demanding to have detections in all the WISE bands we are biasing ourselves to the most extreme IR-bright sources.

In total, we chose 48 observables from \textit{Gaia} DR2, 2MASS, WISE, and IPHAS and VPHAS+ data. The colours used are $r-H\alpha$ plus all combinations of the passbands of \textit{Gaia} DR2, 2MASS, and WISE (i.e. $G_{BP}-G$, $G_{BP}-G_{RP}$, $G_{BP}-J$, $G_{BP}-H$, $G_{BP}-K_{s}$, $G_{BP}-W1$, $G_{BP}-W2$, $G_{BP}-W3$, $G_{BP}-W4$, $G-G_{RP}$, $G-J$, $G-H$, $G-K_{s}$, $G-W1$, $G-W2$, $G-W3$, $G-W4$, $G_{RP}-J$, $G_{RP}-H$, $G_{RP}-K_{s}$, $G_{RP}-W1$, $G_{RP}-W2$, $G_{RP}-W3$, $G_{RP}-W4$, $J-H$, $J-K_{s}$, $J-W1$, $J-W2$, $J-W3$, $J-W4$, $H-K_{s}$, $H-W1$, $H-W2$, $H-W3$, $H-W4$, $K_{s}-W1$, $K_{s}-W2$, $K_{s}-W3$, $K_{s}-W4$, $W1-W2$, $W1-W3$, $W1-W4$, $W2-W3$, $W2-W4$, $W3-W4$).
The idea behind using all these combinations is that we do not know which colours are ideal for selecting PMS objects, so we let principal component analysis (PCA) facilitate this (see Sect. \ref{PCA}). The reason why neither $r$ nor $H\alpha$ passbands are combined with the other passbands is explained in Sect. \ref{T_Herbig}.
 
In addition, we constructed two observables, $G\textsubscript{var}$ and $V\textsubscript{htg}$, that trace optical photometric variability and are based on the \textit{Gaia} passbands. We define $G\textsubscript{var}$ as:

\begin{equation}\label{Gvar}
G\textsubscript{var}=\frac{F'\textsubscript{G}\;e(F\textsubscript{G})\sqrt{N\textsubscript{obs,G}}}{F\textsubscript{G}\;e'(F\textsubscript{G})\sqrt{N'\textsubscript{obs,G}}},
\end{equation}

where $F\textsubscript{G}$ and $e(F\textsubscript{G})$ are the \textit{Gaia} G band flux and its associated uncertainty for a certain source and $N\textsubscript{obs,G}$ the number of times that source was observed in the G band. The idea is that variable sources have larger uncertainties (weighted with the square root of the number of observations) than non-variable ones. $F'\textsubscript{G}/e(F'\textsubscript{G})\sqrt{N'\textsubscript{obs,G}}$ refer to the median value of \textit{Gaia} DR2 sources of the same brightness. This denominator is necessary as non-variable objects of different brightness show different median uncertainties. A similar indicator was used in \citet{2018A&A...620A.128V} to study the variability of known Herbig Ae/Be stars. It was evidenced that this variability proxy mostly traces irregular (i.e. non-periodic) variabilities caused by material on the line of sight, so we expect it to be efficient in separating CBes from HAeBes. We define the heterogeneous variability ($V\textsubscript{htg}$) as:

\begin{equation}\label{Var_col}
V\textsubscript{htg}=\frac{F'\textsubscript{B\textsubscript{p}}\;e(F\textsubscript{B\textsubscript{p}})\sqrt{N\textsubscript{obs,B\textsubscript{p}}}}{F\textsubscript{B\textsubscript{p}}\;e'(F\textsubscript{B\textsubscript{p}})\sqrt{N'\textsubscript{obs,B\textsubscript{p}}}}-\frac{F'\textsubscript{R\textsubscript{p}}\;e(F\textsubscript{R\textsubscript{p}})\sqrt{N\textsubscript{obs,R\textsubscript{p}}}}{F\textsubscript{R\textsubscript{p}}\;e'(F\textsubscript{R\textsubscript{p}})\sqrt{N'\textsubscript{obs,R\textsubscript{p}}}}.
\end{equation}

This $V\textsubscript{htg}$ observable is based on the same idea as $G\textsubscript{var}$ but it evaluates the heterogeneous variability that may be present among the blue ($G_{BP}$) and red ($G_{RP}$) filters. This may arise, for example, by circumstellar material causing irregular extinction episodes (as is the case in the reddening and blueing associated during the variations of UX Ori type stars, \citealp{2000ASPC..219..216G}) or by variable accretion.

\subsection{Features}\label{PCA}
We use PCA to select the optimal set of features for our problem. When applying PCA to our complete set of 48 observables we obtain 48 principal components. However, in our pipeline only 12 of those principal components carry $99.99$\% of the variance (see Appx. \ref{Traing_Test}). These principal components that carry almost all of the variance of the space of observables constitute our set of features. In other words, these principal components are the features used by the ANN. PCA also removes any linear dependency between the observables.

\subsection{Evaluation metrics}\label{S_metrics}

We use two correlated metrics, precision (P) and recall (R). They are defined as follows:

\begin{equation}\label{E_Precision}
P=\frac{TP}{TP+FP},
\end{equation}

where TP is the number of true positives, that is, the number of sources of a certain category correctly catalogued, and FP is the number of false positives, this is the number of sources of the same category wrongly classified\footnote{Other metrics, like the Area Under the Curve or the F\textsubscript{1} score, were discarded because FP is over-measured in this classification problem due to contamination in the training data, see Appx. \ref{Weights}.}. In other words, of all objects for which we have predicted a certain  category, P describes what fraction was correctly classified. Separately:

\begin{equation}\label{E_Recall}
R=\frac{TP}{TP+FN},
\end{equation}

where FN is the number of false negatives, that is, the number of sources that belong to a certain category but were not classified as such. In other words, of all objects that are actually of a certain class, R describes what fraction have we detected as belonging to that class, introducing a notion of completeness. These metrics are defined independently for each category.

\subsection{Data}\label{S_Data}

Before describing the training data, it is necessary to assess how many sources exist with all the observables we are using (Sample of Study, SoSt hereafter). The first step for generating this SoSt was to cross-match the catalogues that contain the required observables (\textit{Gaia} DR2, AllWISE, IPHAS, and VPHAS+). Examples of work where this was done to a high level of accuracy are \citet{2018MNRAS.481.3357S} for \textit{Gaia} DR2 with IPHAS and \citet{2019A&A...621A.144M} and \citet{2018MNRAS.481.2148W} for \textit{Gaia} DR2 with AllWISE, among others. However, these cross-matches arrived at a high level of accuracy by sacrificing completeness. PMS sources in particular, because of their variability and preferred location in extincted and crowded regions,  tend to be excluded in those general cross-matches (e.g. only $\sim 52\%$ of the known HAeBes are present in the AllWISE `BestNeighbour table' of \citealp{2019A&A...621A.144M}). Instead, we perform a more generous cross-match accepting that we may generate some incorrect associations.
 
We first cross-matched \textit{Gaia} DR2 (using epoch 2000 adapted coordinates) with IPHAS and VPHAS+ independently with a 1 arcsecond aperture because that is approximately the angular resolution of VPHAS+, IPHAS being slightly worse. We found that 95\% of the sources are within 0.25 arcsecond. These two catalogues present a further complication. They present different observations of the same source as different entries and hence produce duplications in the cross-match. Therefore, in those cases we chose the observation with data in all the passbands, if any. If none or more than one of the observations have information in all the passbands we chose the one with a higher quality flag and, in the case of having the same flags, we chose the object with the smaller angular distance to the \textit{Gaia} DR2 source. Similarly, whenever a \textit{Gaia} DR2 source was present in both IPHAS and VPHAS+ we gave priority to the observation with all the passbands, followed by the one with a higher quality flag and, in the case of having the same flags, to the object with the smaller angular distance to the \textit{Gaia} DR2 source. Then, we performed another cross-match using \textit{Gaia} DR2 coordinates with AllWISE, using a cross-match aperture of 2 arcsecond. This cross-match aperture, though large, was chosen after the experience in \citet{2018A&A...620A.128V} where even a 3 arcsecond aperture was still not sufficient for some HAeBes. We found that 95\% of the sources are within 1.12 arcsecond. This last cross-match provides us with a set of 51,548,230 sources. However, missing values are not allowed in ANNs and only 4,151,538 sources ($8$\% of the original set) have all the 48 observables (see Sect. \ref{Observables}).  This constitutes our SoSt, the master sample of all the objects with the data necessary to enter the ANN. This set has a mean of $G=16.7\pm2.0$ mag (error is $1\sigma$ of the mean) so 98\% of the sources are in the range $12.3<G<20.3$ mag. The mean parallax is $\varpi=0.36\pm0.75$ mas. We note that the \textit{Gaia} parallax is not available for all the sources. The sky footprint of the SoSt is not homogeneous as it is limited by the combined footprint of the surveys used. Primarily, IPHAS and VPHAS+ are limited to the galactic plane ($5.5^{\circ}>b>-5.5^{\circ}$) and VPHAS+ footprint ($29^{\circ}>l>-145^{\circ}$) is largely incomplete at the time of writing. In addition, spurious WISE photometric detections in the galactic plane are a known issue (\citealp{2019MNRAS.487.2522M} and references therein). Furthermore, due to the \textit{Gaia} scanning law, \textit{Gaia} DR2 itself presents a heterogeneous footprint completeness. Finally, demanding proper detections up to W4 ($22\mu m$) and in the $H\alpha$ passband excludes many objects and it may be expected to have overdensities of SoSt sources around star forming regions. The impact of this footprint in the final catalogues is addressed in Sect. \ref{S_Evaluation}.

\begin{figure*}[ht!]
\includegraphics[scale=0.78]{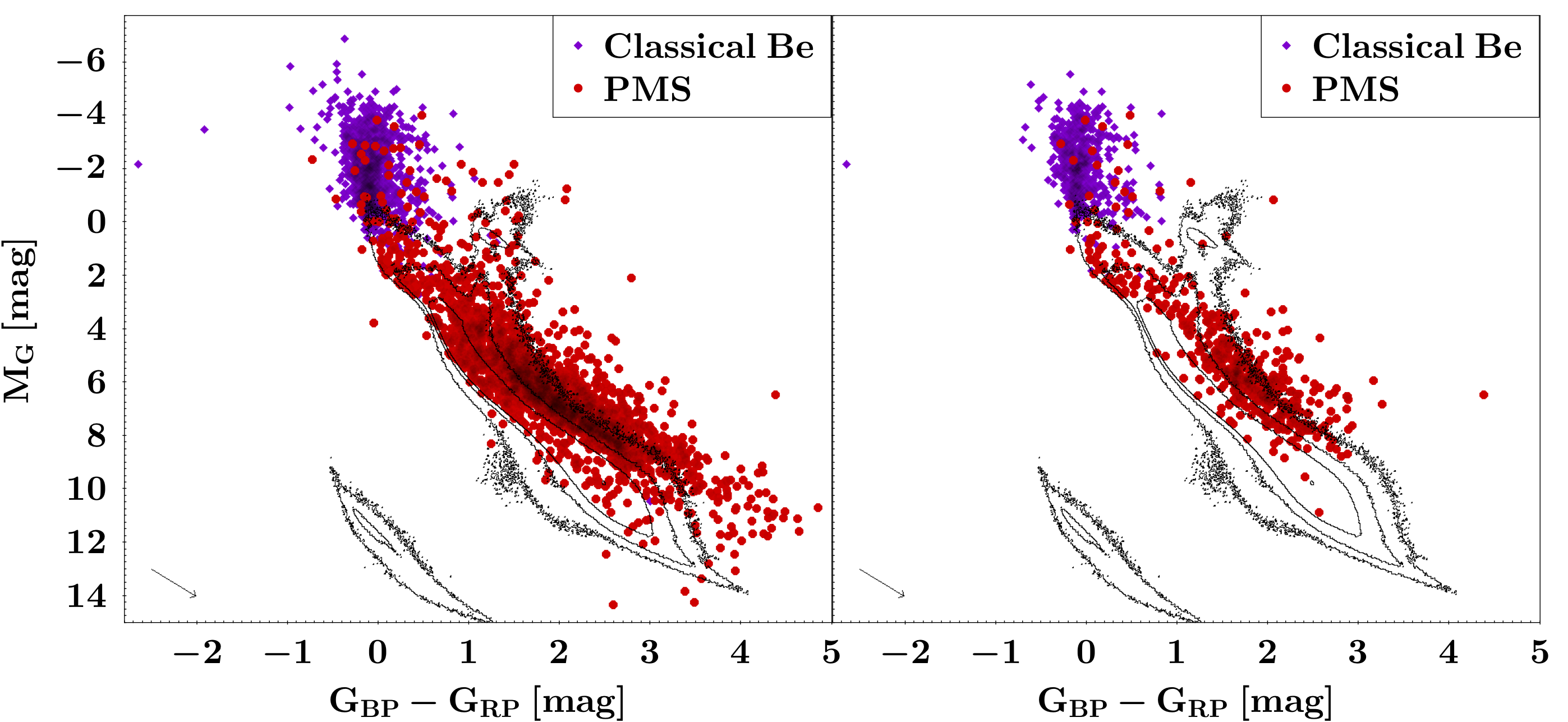}
\caption{\textit{Gaia} colour vs. absolute magnitude diagram. Known PMS (in red circles) and classical Be stars (in violet diamonds) with good astrometric solution and corrected from interstellar extinction are plotted. An extinction vector corresponding to $A_{G}=1$ is shown at the bottom left of each plot. Black contours trace \textit{Gaia} sources within 500 pc with good astrometric solution. \textit{Left:} All known sources. \textit{Right:} The subset of sources with all the observables that are used for training. The very blue classical Be star is \object{$\omega$ CMa} and it probably has a spurious $G_{RP}$ magnitude because of being brighter than the bright limit of \textit{Gaia} DR2.}\label{fig:Train}
\end{figure*}

As the beams of IPHAS, VPHAS+, and AllWISE are larger than \textit{Gaia}'s, different \textit{Gaia} sources could have been assigned to the same IPHAS, VPHAS+, or AllWISE source. This can be the case if various \textit{Gaia} sources are present within the same beam or if a wrong assignation was done in our generous cross-match. Indeed, $4.9\%$ of the AllWISE sources are repeated and $0.31\%$ of the IPHAS or VPHAS+ objects. These do not affect the classification, as the values are too small to have a significant impact on the training or the final catalogues (see Sects. \ref{Other_sources} and \ref{S_Evaluation} respectively). However, this implies that on average 1/42 (regarding AllWISE) and 1/625 (regarding IPHAS or VPHAS+) sources of the SoSt are fake, in the sense that its associated photometry does not belong to them, or it is a mixture of all the \textit{Gaia} sources within the same beam. Another way of estimating the number of purely incorrect cross-matches is by comparing the \textit{Gaia} passbands and colours with the AllWISE and IPHAS or VPHAS+ ones. In the case of AllWISE we compared $G_{RP}-J$ vs. $J-H$, which are strongly linearly correlated, and labelled as potential incorrect cross-matches those sources that were beyond 0.5 mag of the best linear fit. This results in about $2.3\%$ bad matches for AllWISE. In the case of H$\alpha$ we compared $G_{BP}$ vs. $r$ (there is no linear relation between colours) and labelled as potential incorrect matches those sources that were beyond 1 mag (to account for variability) of the best linear fit. This results in a contamination of $1.3\%$ for IPHAS or VPHAS+. Therefore, we conclude that the cross-matches are good to the $\sim 98\%$ level.

We did not take into account the quality flags of the catalogues. This decision was made for two reasons. First, IPHAS and VPHAS+ have very stringent quality indicators, and by limiting ourselves to sources with a good flag in these catalogues we reduce the size of our training set significantly (e.g. the SoSt would be reduced to $47\%$). Similarly, the mid-IR colours W3 and W4 tend to have very poor quality flags. Only $\sim 10\%$ of the sources within \textit{Gaia} and AllWISE with information in all passbands have reliable mid-IR measurements (\citealp{2019MNRAS.487.2522M}). However, in this work the mid-IR is of paramount importance and cannot be excluded, as it is where the discs around HAeBes start to differ from the dust-free discs around classical Be stars (\citealp{1988A&A...203..348W}; \citealp{2013A&ARv..21...69R}). Second, because introducing cuts in the training data based upon quality criteria can introduce uncontrolled biases in the subsequent selection. This is because these quality flags are a result of a combination of very different factors. It is preferable to let the ANN deal with bad quality photometry as well as contaminants. Nonetheless, these quality flags are added to the final catalogues of new PMS and CBe candidates (Sect. \ref{S_out} and Tables \ref{Table2} and \ref{Table3}). The consequences of using low-quality data are discussed in Sect. \ref{S_Evaluation}.

\section{Labelled sources}\label{Trainingset}

As described in Sect. \ref{S_Introduction}, we need to select which categories we want the ANN to learn how to classify. Then we need to label a set of sources as belonging to these categories and use them to train the ANN. These labels are considered as ground truth and any bias, trend or contamination of this sample inevitably results in a bias in the final classification. In this section we describe the construction of this set of Labelled Sources, which is a subset of the Sample of Study. The complementary subset of the SoSt that is not labelled (Input Set) is the one classified by the trained ANN (see Appx. \ref{S_Description of the pipeline} and Fig. \ref{fig:Pipeline2} for further details).

We use one category of PMS sources and another category of classical Be stars, as telling the difference between these two groups is the main goal of the algorithm. In addition of learning from the characteristics of PMS and CBe objects we need the algorithm to learn from the existence of other similar or distinct sources that do not belong to these categories. This includes the erroneous or spurious data present in every catalogue. In other words, we need to construct a representation of what the algorithm finds when being applied to the Input Set. Hence, we use a third category of other objects, which comprises all types of sources present within the catalogues used that are neither a PMS source nor a CBe star. Therefore, the set of Labelled Sources contains already known PMS sources (Sect. \ref{PMS}), already known CBes (Sect. \ref{Classic Be stars}), and other objects (Sect. \ref{Other_sources}). In the following sections the construction of these three categories is described.

All known PMS and CBe sources considered with a good astrometric solution appear on the \textit{Gaia} HR diagram (colour vs. absolute magnitude diagram) in Fig. \ref{fig:Train}. We define as sources with a `good astrometric solution' those with a re-normalised unit weight error (RUWE parameter of \textit{Gaia} DR2) of smaller than 1.4 and $\varpi/\sigma(\varpi)\geq10$. Only these astrometrically well behaved sources have trustworthy positions in the HR diagram, as astrometry carries most of the uncertainty (see e.g. \citealp{2018A&A...620A.128V}). However, those with a bad astrometric solution are still used by the algorithm as the observables are astrometry-independent (see Sect. \ref{Observables}). In this work we use the parallax to distance conversion of \citet{2018AJ....156...58B}. In order to achieve the most accurate HR diagram positions we also needed to correct for extinction. Unfortunately, it cannot be totally taken into account as in general the intrinsic extinctions are unknown. However, we corrected for interstellar extinction by using the dust map of \citet{2019A&A...625A.135L} and the extinction coefficients of \textit{Gaia} of \citet{2018MNRAS.479L.102C}. This interstellar extinction shall only be considered as a lower limit to the total extinction. This procedure for generating HR diagrams is standard throughout the paper, so all the HR diagrams presented can be directly compared.

\subsection{PMS object category}\label{PMS}
Although for the algorithm there is just a single class of PMS objects, we create that class by combining intermediate mass Herbig Ae/Be stars and lower mass T-Tauris, so we cover the whole mass range. 

\subsubsection{Herbig Ae/Be stars}\label{T_Herbig}
Regarding the Herbig Ae/Be stars, we start with the compilation of \citet{2018A&A...620A.128V} where most known HAeBes could be matched with \textit{Gaia} DR2 data. The main issue with Herbig Ae/Be stars is that almost all of them are brighter than the bright limit of IPHAS and VPHAS+ ($12-13$ mag). Using H$\alpha$ equivalent widths (EWs) we derived the IPHAS and VPHAS+ like colour $r-H\alpha$ using the synthetic tracks of \citet[see their Fig. 6, extinctions and effective temperatures are present in \citealp{2018A&A...620A.128V} and references therein]{2005MNRAS.362..753D}. Combining the H$\alpha$ EWs of \citet{2018A&A...620A.128V} and \citet{2020MNRAS.493..234W} with the few sources present in IPHAS or VPHAS+ gave us $r-H\alpha$ colour for 215 HAeBes. This is why neither $r$ nor $H\alpha$ passbands are combined with the rest in Sect. \ref{Observables}, as we do not have them for many sources. There is a bias in this conversion from H$\alpha$ EWs to $r-H\alpha$ colour because it can only be applied to those objects with observed H$\alpha$ in emission above the continuum. Hence, it could not be applied to the many HAeBes with intrinsic emission filling in the underlying absorption but below the continuum level. This bias also appears later for T-Tauri stars and CBes in Sects. \ref{TTauri} and \ref{Classic Be stars} and its impact is addressed in Sect. \ref{S_Evaluation}.

The cross-match with AllWISE to obtain 2MASS and WISE passbands was already performed in \citet{2018A&A...620A.128V}. The final number of Herbig Ae/Be stars considered is 255, of which 163 have all observables. We did not include Massive Young Stellar Objects (\citealp{2013ApJS..208...11L}) in this sample as in general they are not optically visible so they are not present in \textit{Gaia} DR2 (except those that are already in \citealp{2018A&A...620A.128V} list which were included in this study). 

\subsubsection{T-Tauri stars}\label{TTauri}
To the set of intermediate mass Herbig Ae/Be stars we add a set of T-Tauri stars to complete the low-mass regime. If we use those objects catalogued as T-Tauris in the SIMBAD database (around 3500 objects at the time of writing) we end up, after the cross-matches, with most of the objects having being catalogued by a few papers dedicated to very specific regions (e.g. \citealp{2014A&A...570A..82V} on NGC 2264 open cluster or \citealp{2013A&A...559A...3S} on Tr 37). In order to minimise the possible implications due to this we add the sources of the Herbig-Bell (HB) Catalogue (\citealp{1988cels.book.....H}) which, although focused in the Orion region, has sources distributed all over the sky. We cross-matched the set of T-Tauris with \textit{Gaia} DR2 with a 0.5 arcsecond aperture (close to the 0.4 arcsecond angular resolution of \textit{Gaia} DR2). We double checked that the cross-matched sources have a similar V and G band (within $\pm2$ mag, the range is rather generous to avoid biasing to exclude very variable sources) when possible to discard bad cross-matches. Then, we cross-matched the \textit{Gaia} source identifications with those of the SoSt (see Sect. \ref{S_Data}) to obtain the T-Tauri stars with all the observables. 

In addition, the HB catalogue provides us with H$\alpha$ EWs and spectral types that allow us to derive $r-H\alpha$ colour for 297 more T-Tauris. To this end, we used the HB B-V colour, which come from simultaneous passbands at maximum brightness, and the spectral types provided by the HB catalogue to derive extinctions for these T-Tauris. Whenever B-V colours were not available we used those of the APASS survey (\citealp{2018AAS...23222306H}) with a 3 arcsecond cross-match. A small error is introduced for objects colder than roughly a G2 V star which are typically given slightly smaller $r-H\alpha$ magnitudes than those that correspond to them (see \citealp{2005MNRAS.362..753D} for further details). The overall result is a sample of 3171 T-Tauri stars, of which 685 have information in all the observables.

\subsection{Classical Be stars}\label{Classic Be stars}
For the classical Be stars, we use the Be Star Spectra Database (BeSS Database, \citealp{2011AJ....142..149N}) which comprises 2264 CBes. This includes the candidates of \citet{2013MNRAS.430.2169R, 2015MNRAS.446..274R}. To these we add 35 more CBes from \citet[those they claim as secure detections]{2018A&A...609A.108S}. We cross-matched that catalogue with \textit{Gaia} DR2 using a 0.5 arcsecond aperture. Again, we double checked that the cross-matched sources have a similar V and G band photometry (within $\pm2$ mag) when possible, in order to discard bad cross-matches. Then, we cross-matched the \textit{Gaia} source identifications with those in the SoSt (Sect. \ref{S_Data}) to assess how many CBe stars are there with all the observables.

In order to increase the number of stars in this category, we complemented it with H$\alpha$ EWs from the spectra available in the BeSS database. We estimated an uncertainty measuring EWs of 15\%, which is probably within the intrinsic EW variations of these objects. Then, we used again the synthetic tracks of \citet{2005MNRAS.362..753D} to transform H$\alpha$ EWs to IPHAS and VPHAS+ $r-H\alpha$ colour for 442 sources. In order to do this, we used the spectral types of the BeSS database to estimate effective temperatures and, if undetermined, we estimated them from the positions in the HR diagram (Fig. \ref{fig:Train}). We assumed no extinction, which is roughly safe for this kind of object (only the faint ones from \citealp{2015MNRAS.446..274R} suffer significantly from interstellar extinction). To assess whether this is a valid assumption we studied the extinction in the G band provided by \textit{Gaia} DR2 for all the CBe stars for which it is available. If we take the central values we found that $94$\% of the sources have an $A_{G}$ lower than $1.55$, which is roughly the value beyond where the extinction becomes significant for the colour conversion of \citet{2005MNRAS.362..753D}. The final number of classical Be stars considered is 1992 of which 775 have information in all the observables. 

\subsection{Disentangling Herbig Ae/Be, CBe stars, and B[e] stars}\label{Disentangling}

There is some inevitable contamination between categories. For example, the set of known PMS objects is contaminated in its massive end by classical Be stars and vice-versa. Indeed, there were 15 sources that appeared both as PMS and CBe star in the previous selections. Therefore, we needed to take decisions on how to catalogue them, even though in many cases there is no clear answer in the literature. We did not exclude these objects as they are the most interesting ones for the algorithm to learn from. The particular cases and the decision made upon them are detailed in Appx. \ref{AppendixA}. 

In addition, within our sets of known sources there were many `unclassified B[e]' stars (also known as FS CMa stars, \citealp{2017ASPC..508..285M}; \citealp{2018PASP..130k4201A}). FS CMa objects are an inhomogeneous group of B stars with forbidden lines and a very unclear nature. These forbidden lines and the dust-type infrared excess exclude them from being PMS or CBe sources (\citealp{2013A&ARv..21...69R}) and we removed them from the sets of known objects in order to not bias the results. The 17 excluded FS CMa sources are also detailed in Appx. \ref{AppendixA}. As a word of caution, independently, around half of the HAeBes display the B[e] phenomenon (see \citealp{2017ASPC..508..175O}).

\begin{figure}[ht!]
\includegraphics[scale=0.713]{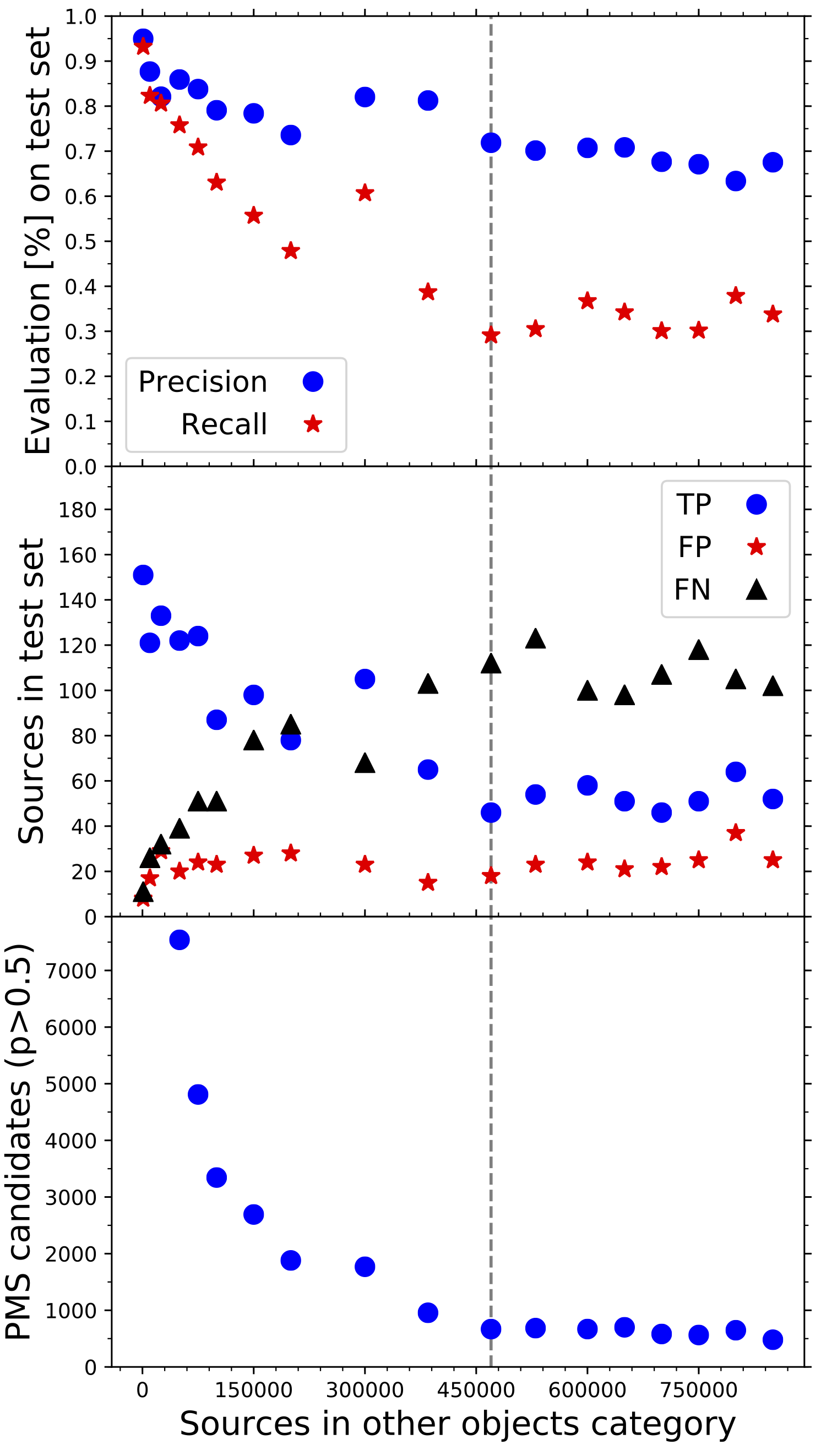}
\caption{Different metrics of the ANN on the PMS category vs. different sizes of the other objects category. Stabilisation point is marked with a vertical grey line. \textit{Top:} Precision and Recall. As the size of the category gets larger the recall drops drastically up to the stabilisation point whereas the precision is roughly stable at all sizes. \textit{Middle:} TP, FP, and FN. Similarly, TP and FN have equal stabilisation point whereas FP is stable for all sizes. \textit{Bottom:} Number of PMS candidates obtained when generalizing the trained ANN, we note the same stabilisation point.}\label{fig:size_other}
\end{figure} 

\subsection{Other objects}\label{Other_sources}

We construct the category of other objects by randomly sampling sources from the Sample of Study. We would like to have a representative set of whatever else might be present in the SoSt that is not a PMS object or a classical Be star. The question is how large this category should be in order for the algorithm to generalise properly. In other words, we want to know how many random sources from the SoSt are necessary so all populations present in the cross-matched catalogues have been represented in this category of other objects. 

This can be estimated by training an ANN with different sizes of this third category, and studying how well it generalises in each case. The size of the previously described categories is kept constant. Using an ANN (3 fully connected hidden layers of 300 neurons each) we evaluate how the precision and recall of the network on the PMS group behave on a test set (sized $20\%$ of the training set) for different sizes of this other objects category (Fig. \ref{fig:size_other}). The architecture of this ANN is a bit more sophisticated than the complexity demanded by our problem (as can be seen by the chosen architecture in Appx. \ref{architecture}), but we wanted to be sure to not underfit in any case so the ANN is always sensitive to new data. If the category of other objects is very small the algorithm is very precise and has a high recall (Fig. \ref{fig:size_other} on top); few other objects appear in the regions of the feature space where PMS and CBe stars tend to be placed, so they have little impact in the classification (it also indicates that the ANN is good in telling the difference between PMS and CBe stars, although we note that a large fraction of the PMS category are low-mass T-Tauri stars). The more we populate the feature space with other objects the algorithm is less able to recognise PMS stars (the false negatives rise, Fig. \ref{fig:size_other} at middle) as the regions of the feature space with the more common PMS sources start to be highly populated by objects similar to PMS stars and undiscovered PMS objects. The number of false positives stays the same as the algorithm is still being efficient in the less populated regions. In other words, the PMS candidate region in the feature space gets smaller and localised around the less common PMS sources. The number of true positives drops as a consequence of the increase of false negatives. This causes the precision to barely change (Eq. \ref{E_Precision}) but the recall to drop (Eq. \ref{E_Recall}, Fig. \ref{fig:size_other} on top) up to a stabilisation point (grey line in Fig. \ref{fig:size_other}) where most of the different types of objects that populate the feature space differently have appeared, and hence adding more sources does not further constrain significantly the locus of PMS candidates in the feature space. 

This stabilisation point can also be found if we study the number of PMS candidates retrieved after generalizing the trained ANN to the unlabelled sources of the SoSt (Fig. \ref{fig:size_other} at bottom, selecting as candidates those with a probability $p\geq50\%$ of belonging to the PMS category). This number drops quickly from a very high value, where the algorithm does not know about the existence of anything but PMS stars and CBes, up to the stabilisation point where it becomes roughly stable. Of course, adding more other objects always diminish the region of PMS and CBe candidates in the feature space, but this stabilisation point constitutes the optimal size for the category of other objects, as larger sizes do not compensate the amount of extra information for the contamination they introduce. 

Therefore, for constructing  the category of other objects, we randomly sample sources from the SoSt (excluding the sources in the PMS and CBe categories) so that they are in a proportion of 99.82\% with the number of sources in the category of PMS objects (848), this being the observed stabilisation point. This scales to 470,263 objects. Some of these sources might have been classified previously by different catalogues, although most remain unclassified.

We can approximate what is the proportion of other objects to PMS objects in the SoSt from a simulation. \citet{2012A&A...543A.100R}, using the \textit{Gaia} Universe Model Snapshot (GUMS) simulation estimate that the percentage of PMS objects within $G<20$ mag in \textit{Gaia} is $0.18$\%. The real proportion of PMS sources in the SoSt is somewhat larger as we are demanding detections up to $22\mu m$ and in $H\alpha$. This implies, theoretically, that roughly there are as many undiscovered PMS stars in the other objects category as known PMS stars in the PMS category.

\begin{figure}[ht!]
\includegraphics[scale=0.416]{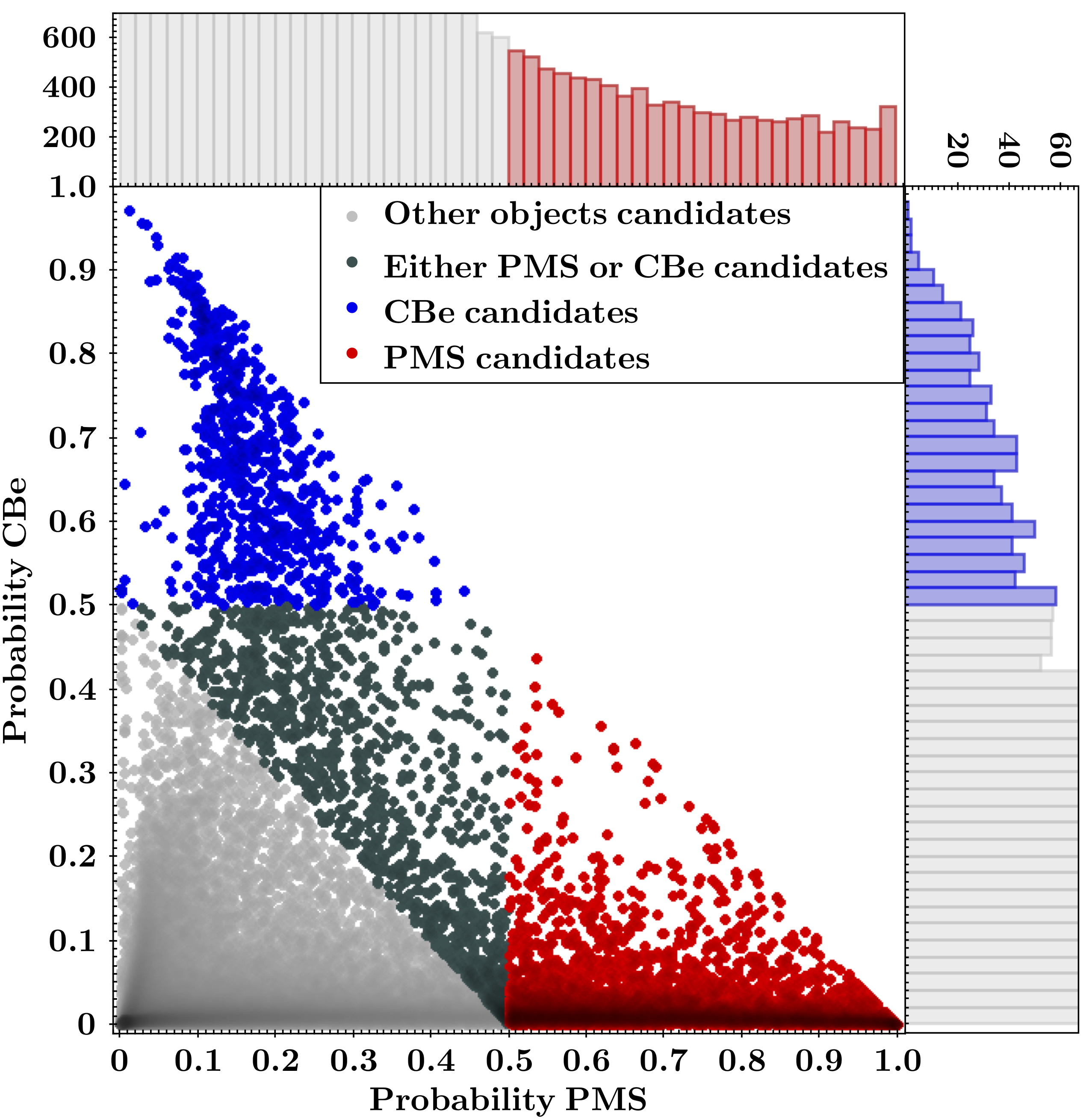}
\caption{Output probability map of the Sample of Study. A probability threshold of $p\geq50\%$ is used to select the PMS (in red) and classical Be candidates (in blue). On top and right number histograms of the candidates for different probabilities. In dark grey the sources which belong to either category ($p(PMS)+p(CBe)\geq50\%$ but $p(PMS)<50\%$, $p(CBe)<50\%$). Uncertainties are not indicated for clarity. Numbers are: PMS candidates (8,470), classical Be candidates (693), either (1,309), other objects (4,140,511). }\label{fig:Probability_map}
\end{figure}

\section{Results}\label{S_out}

The set of Labelled Sources described in the previous section and the Sample of Study of Sect. \ref{S_Data} are introduced in a pipeline of algorithms which core is an ANN. As an output, the pipeline gives probabilities and associated uncertainties for all the sources in the SoSt of belonging to either the PMS, classical Be, or other objects category (excluding the known PMS and CBe sources included in the categories of the Labelled Sources set). These probabilities, that sum up to one in each source, are presented in Fig. \ref{fig:Probability_map}. The pipeline, algorithms, and methodology are described in detail in Appx. \ref{S_Description of the pipeline}.

The SoSt with probabilities is available in electronic form in its entirety (4,150,983 sources). This data is made available so the user can choose their own probability threshold ($p$) to select PMS and classical Be candidates. Choosing $p$ implies fixing a precision (P) and a recall (R). The pipeline also gives a solid estimate of the precisions and recalls for different $p$ thresholds. However, due to the nature of the pipeline the values for the precision are only lower limits (see Appx. \ref{Weights}). Ideally these two metrics should be as high as possible but there is a trade-off between them. This is shown in Fig.~\ref{PrevsRe}, where the precision and recall for both the PMS and CBe category are plotted for different $p$ probability thresholds. Rising the threshold to $p\geq99\%$ maximises the precision to almost 1, but as a consequence the recall lowers to almost 0. The opposite also applies and neither of both extremes is close to be representative of a good selection; as it would be either largely incomplete or largely imprecise. The general shape of the curves is determined by the architecture of the algorithm and the peculiarities of the classification problem (see Appx. \ref{architecture}). 

\begin{figure*}[ht!]
\includegraphics[scale=0.39]{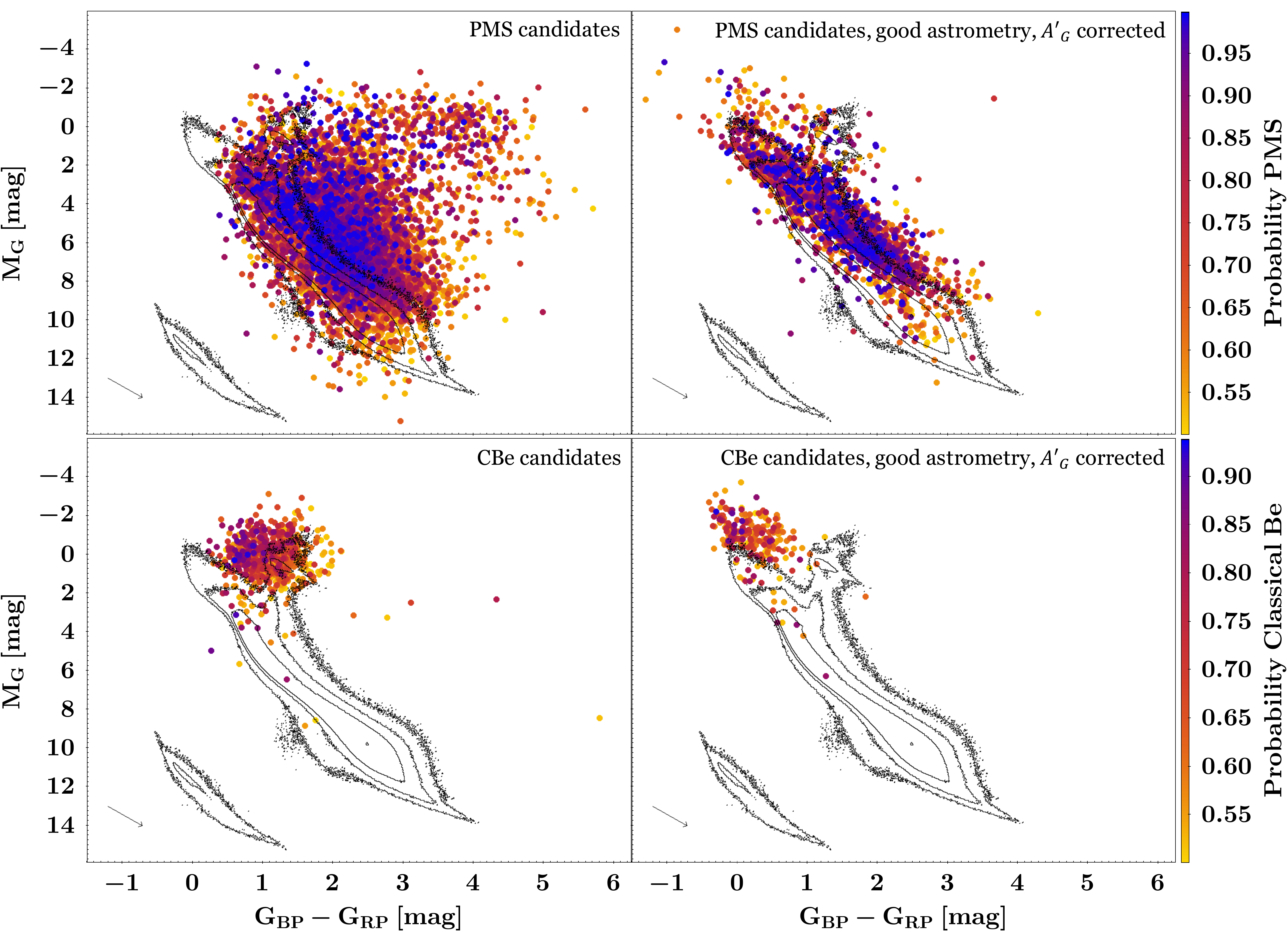}
\caption{\textit{Gaia} colour vs. absolute magnitude HR diagram. An extinction vector corresponding to $A_{G}=1$ is shown on the bottom left of each plot. Black contours trace \textit{Gaia} sources within 500 pc with good astrometric solution. \textit{Top left:} HR diagram of PMS candidates ($p\geq50\%$) colour-coded by their associated membership probability. \textit{Top right:} HR diagram of PMS candidates ($p\geq50\%$) with good astrometric solution colour-coded by their associated membership probability and corrected from interstellar extinction. \textit{Bottom left:} HR diagram of classical Be candidates ($p\geq50\%$) colour-coded by their associated membership probability. \textit{Bottom right:} HR diagram of classical Be candidates ($p\geq50\%$) with good astrometric solution colour-coded by their associated membership probability and corrected from interstellar extinction. \label{HR_grid}}
\end{figure*}

In practice, using probability thresholds below 50\% is possible, but entering the regime where the algorithm assigns larger probabilities to other categories is not advisable as $p$ does not correlate linearly with the precision and recall (see Fig. \ref{PrevsRe}). At $p\geq50\%$ the resulting catalogues are: new PMS candidates (8,470 sources, $P=40.7\pm1.5\%$, $R=78.8\pm1.4\%$), new classical Be candidates (693 sources, $P=88.6\pm1.1\%$, $R=85.5\pm1.2\%$). We note that the precisions are lower limits. These catalogues of new candidates are presented in Appx. \ref{AppendixD} in Tables \ref{Table2} and \ref{Table3} respectively and highlighted in Fig. \ref{fig:Probability_map} (full tables available in electronic form)\footnote{As a word of caution, these recalls do not imply that the presented catalogues contain $\sim80\%$ of the existing PMS and CBe stars within any region. They imply that $\sim80\%$ of the known PMS and CBe stars in the test set are recovered by the algorithm. This is for example affected by what the different surveys used are probing and the distribution of the SoSt (see Sect. \ref{S_Data}). As explained in Sect. \ref{Quality assessment}, probably some of the less extreme objects in the observables used have not been classified. Similar reasoning can be applied to the precision values.}. In those tables, together with the probabilities we present the observables used for the training (Sect. \ref{Observables}) and \textit{Gaia} astrometric information. In addition, we included the derived interstellar extinction ($A_{G}'$) and $A_{G}'$ corrected $M_{G}$ and $G_{BP}-G_{RP}$ for those sources with RUWE<1.4 and $\varpi/\sigma(\varpi)\geq5$. These allow for a better positioning in the HR diagram (see Sect. \ref{Trainingset} and Figs. \ref{fig:Train}, \ref{HR_grid}, and \ref{Progress}). 

In the CBe case the precision does not drop drastically (see Fig. \ref{PrevsRe}). This implies that for the algorithm it is easier to find CBe stars than PMS stars as their locus in the feature space is less prone to contaminants but mostly because there are fewer unclassified CBes in the SoSt (see Appx. \ref{Weights}). A consequence of this is that we retrieve an order of magnitude less CBe candidates than PMS candidates.

Following the discussion of Sect. \ref{Other_sources}, the size of the other objects category roughly coincides with the point where there is approximately one undiscovered PMS source per known PMS source in the training and test sets. Taking this into account, the lower limit on the precision of $P\sim 40\%$ for the PMS group obtained with $p\geq50\%$ is an adequate enough result (i.e. the real precision is roughly double). However, as the precision is a lower limit, it is hard to assess whether a higher probability threshold is better to retrieve a stronger catalogue of PMS candidates. In order to decide this, we need to use parameters and observables that have not been used in the training, and are hence independent of the selection. As explained in Sect. \ref{S_parameters}, the set of features (and hence the classification) is distance and position independent, at least at first order. This means that we can use the HR diagram and the sky locations to assess this issue. 

Before analyzing these catalogues, we first remove the sources brighter than the typical bright limit of IPHAS and VPHAS+ that show significant differences between their IPHAS or VPHAS+ magnitudes and their \textit{Gaia} magnitudes (marked in Tables \ref{Table2} and \ref{Table3} with a `X-mtch' flag). These objects did not affect the training as they barely account for $0.5\%$ of the other objects category. There are 18 PMS and 57 CBe candidates with this flag. These sources are likely to be incorrect cross-matches and they are left out in the following analyses.

\subsection{Evaluation using the HR diagram}\label{Evaluation on HR}

\begin{figure*}[ht!]
\centering\includegraphics[scale=0.5639]{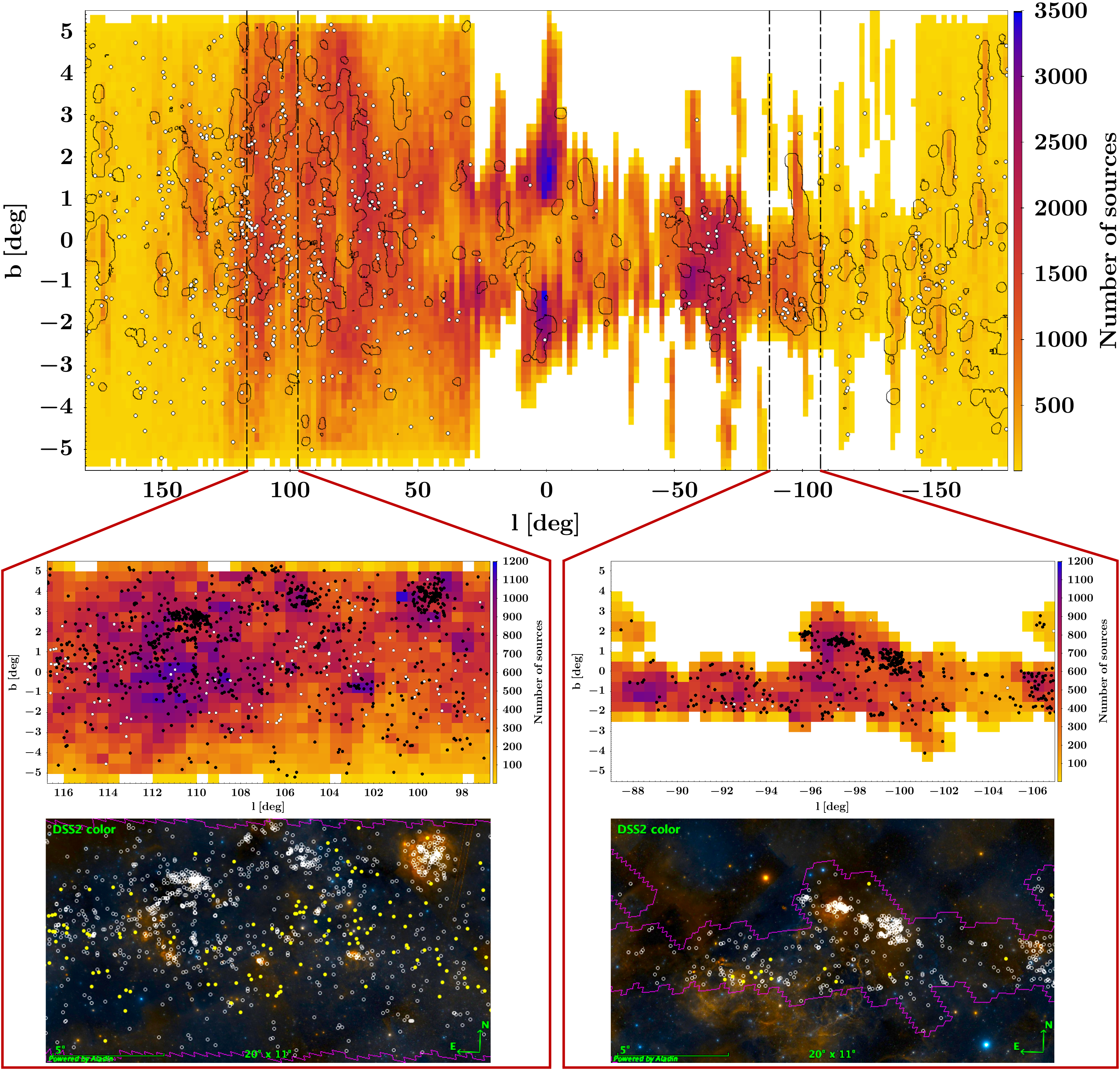}
\caption{\textit{Top:} Sky footprint of the Sample of Study in galactic coordinates, colour-coded by number density. We note the heterogeneity of the footprint. The scarcity of sources between $29^{\circ}>l>-145^{\circ}$ is due to the incompleteness of VPHAS+ at the time of writing. Each pixel is $2 ^{\circ}\times0.2 ^{\circ}$. PMS candidates overdensities appear as black contours. There are ten time more candidates inside than outside the contours. CBe candidates appear as white dots. Expanded regions at bottom panels appear between dashed lines. \textit{Middle:} Expanded regions. Each region is $20^{\circ}\times11^{\circ}$. PMS contours are replaced by PMS candidates (black dots). Each pixel is $0.5^{\circ}\times0.5^{\circ}$. \textit{Bottom:} Same expanded regions in DSS2 colour with the PMS candidates as white circles and the CBe candidates as yellow dots. Contours trace the footprint of the Sample of Study.}\label{Closer_look.pdf}
\end{figure*} 

The HR diagram is not entirely selection independent, as we used different colours in the classification and we do not correct for the unknown intrinsic extinctions. However, the location in the HR diagram, which carries information about evolutionary status, is almost independent of the classification.

The \textit{Gaia} HR diagram of the PMS and classical Be candidates (those with $p>50\%$) can be seen in Fig. \ref{HR_grid} at left panels. In Fig. \ref{HR_grid} at right panels we also distinguish those with a good astrometric solution (RUWE<1.4 and $\varpi/\sigma(\varpi)\geq10$) to which interstellar extinction corrections have been applied. These well behaved sources have trustworthy positions in the HR diagram. The candidates have been colour-coded according to their membership probability to the corresponding category. In this HR diagram we can evaluate the quality of the retrieved catalogues. Regarding the PMS candidates, the majority are placed to the right of the main sequence, as expected for PMS sources. Moreover, if we move up to higher probability thresholds or to those sources with a good astrometric solution we constrain the selection of PMS candidates to those sources that are located in the more likely PMS positions. Something similar happens with the CBe candidates, as they are placed where CBe stars are supposed to be. This can be better appreciated when comparing these candidates with the locus of the known PMS and classical Be stars in Fig. \ref{fig:Train}. We note that $\sim 6\%$ of PMS candidates and $\sim 1\%$ of CBe candidates lack parallax information.

This, in addition of evidencing the quality of the selection, allows to select a higher probability threshold by looking at the retrieved candidates in the HR diagram of astrometrically well behaved sources corrected from interstellar extinction (Fig. \ref{HR_grid} at top right). This threshold can be adapted to the requirements of different studies or situations. Here, we stick to the probability threshold of $p\geq50\%$ for constructing the catalogues of candidates. This is because the candidates with the higher probabilities are the easier ones to find. Hence, as can be seen in Fig. \ref{HR_grid}, most of the high-mass PMS candidates do not have high associated probabilities as the algorithm struggles more to differentiate them from classical Be stars and vice-versa. In addition, Fig. \ref{fig:Probability_map} shows that there are very few CBe candidates with a negligible PMS probability. Therefore, a more conservative selection of the probability threshold would exclude many of the high-mass objects (see histograms of Fig. \ref{fig:Probability_map}).

Finally, although in the rest of the paper we discuss the catalogues as of $p\geq50\%$, the user can construct their own catalogues by means of Tables \ref{Table2}, \ref{Table3}, and the SoSt with probabilities. As any new catalogue is likely to have a higher probability threshold, the discussion and analysis that follows holds true. Hence, from here onward we refer to PMS and CBe candidates as those with a $p>50\%$ in their respective categories.

\begin{figure*}[ht!]
\centering\includegraphics[scale=0.662]{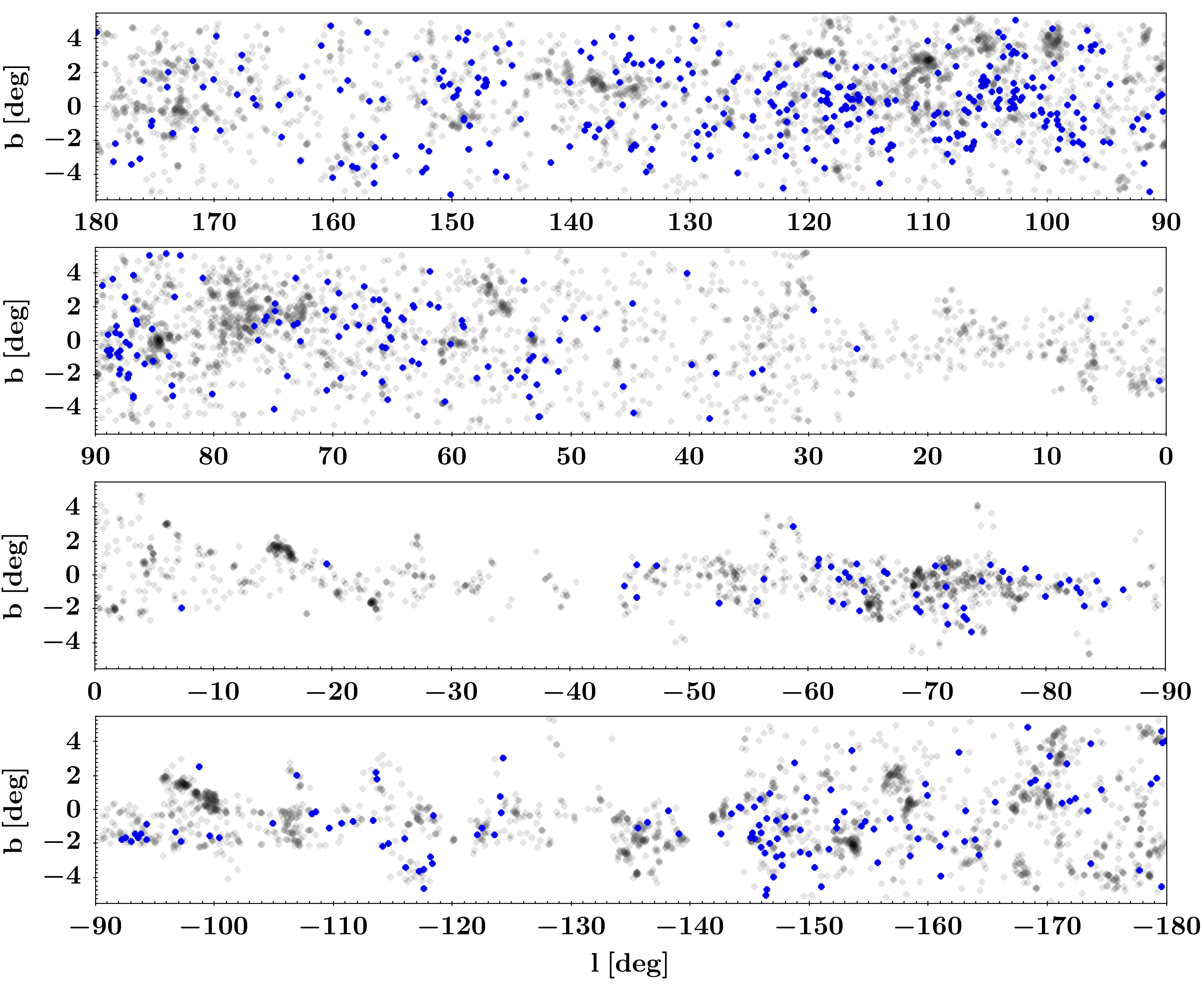}
\caption{Classical Be candidates in blue distributed in the sky in galactic coordinates plotted on top of the PMS candidates in light grey. The densest regions of PMS candidates appear darker. The scarcity of sources between $29^{\circ}>l>-145^{\circ}$ is due to the incompleteness of VPHAS+ at the time of writing.}\label{Sky_Be}
\end{figure*} 

\subsection{Evaluation using sky locations}\label{S_Sky}

The selection has been totally independent of coordinates. This, though true, is limited by the combined footprint of the surveys used; for example, it is limited to the Galactic plane, see Sect. \ref{S_Data}. Now, in Fig. \ref{Closer_look.pdf} at top, we plot the contours of the catalogue of new PMS candidates on the Sample of Study footprint. The PMS candidates trace some of the overdensities of the SoSt. This is because any random selection of sources traces the footprint of the SoSt but might also be because the new PMS candidates are mostly appearing in star forming regions, which would be a strong assessment of the selection.

\begin{figure*}[ht!]
\centering
\includegraphics[scale=0.95]{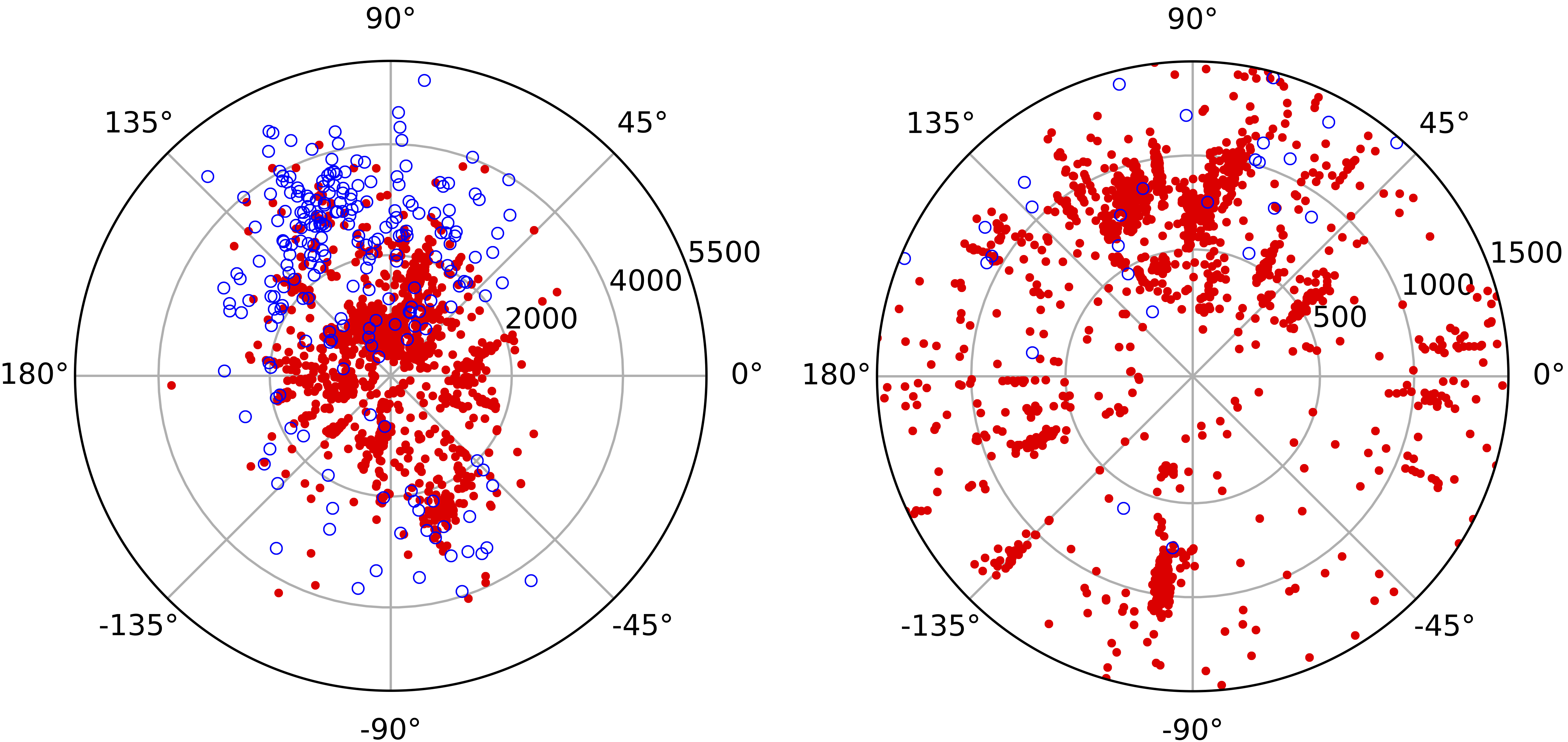}
\caption{Galactic longitude vs. distance (in parsec [pc], from \citealp{2018AJ....156...58B}) of PMS (red dots) and classical Be (blue circles) candidates with good astrometric solution. \textit{Left:}All candidates, \textit{Right:} Candidates up to 1500 pc.\label{Sky_dist}}
\end{figure*}

In this respect, it is noticeable that some overdensities are not particularly populated by PMS candidates. Moreover, if we look at the small scale (examples in Fig. \ref{Closer_look.pdf} at middle and bottom panels) we see that the PMS candidates are not strictly following the SoSt overdensities but are more likely associated to nebulosities. In addition, in Fig. \ref{Sky_Be} we plot all the PMS candidates in the sky. They appear distributed all over the Galactic plane but there are associations of candidates, regions of around $\sim 0.5$ to $1$ squared degrees where there are ten to a hundred more PMS candidates than normally distributed. These associations also appear if we include the distances (Fig. \ref{Sky_dist}). 

This means that the \textit{Gaia} coordinates are assessing the efficiency of the algorithm, as the retrieved PMS candidates are prone to appear around nebulosities and star forming regions, even though these regions are not over-represented in the input data. Further evidence in this respect can be seen in Fig. \ref{Sky_Be}, where the new classical Be candidates are also plotted in the sky. These candidates are distributed all over the Galactic plane but they are not tracing the associations of PMS candidates or nebulosities (see e.g. Fig. \ref{Closer_look.pdf} at bottom panels), which implies that they are indeed of an independent nature for the algorithm. Moreover, if we include the distances (Fig. \ref{Sky_dist}), CBe candidates also appear decoupled from the PMS candidates. CBe candidates are typically further away, something expected from bright B-type stars.

Although the clustering properties of this new set of PMS sources is beyond the scope of this study, we can make some remarks. Firstly, on a global scale, PMS candidates trace some of the regions with more data available. This is likely because these zones contain star forming regions, as not all the regions with more data available are overpopulated with PMS candidates (Fig. \ref{Closer_look.pdf} at top panel). However, on a local scale, PMS candidates do not trace overdensities in the available data, and the associations of candidates appearing are normally not related to those overdensities (e.g. Fig. \ref{Closer_look.pdf} at middle panels). Secondly, the PMS candidates seem to trace nebulosities, and the large sky associations obtained are mostly related to them (e.g. Fig. \ref{Closer_look.pdf} at bottom panels). There are also a few smaller associations of PMS candidates unrelated to footprint overdensities that seem to trace dark nebula and are placed on their edges. Lastly, among the PMS candidates there is no significant correlation between PMS probability ($0.5\geq p\geq1$, see Fig. \ref{fig:Probability_map}) and coordinates.

\subsection{Herbig Ae/Be candidates}\label{Herbig_candidate}

\begin{figure}[t]
\includegraphics[scale=0.28]{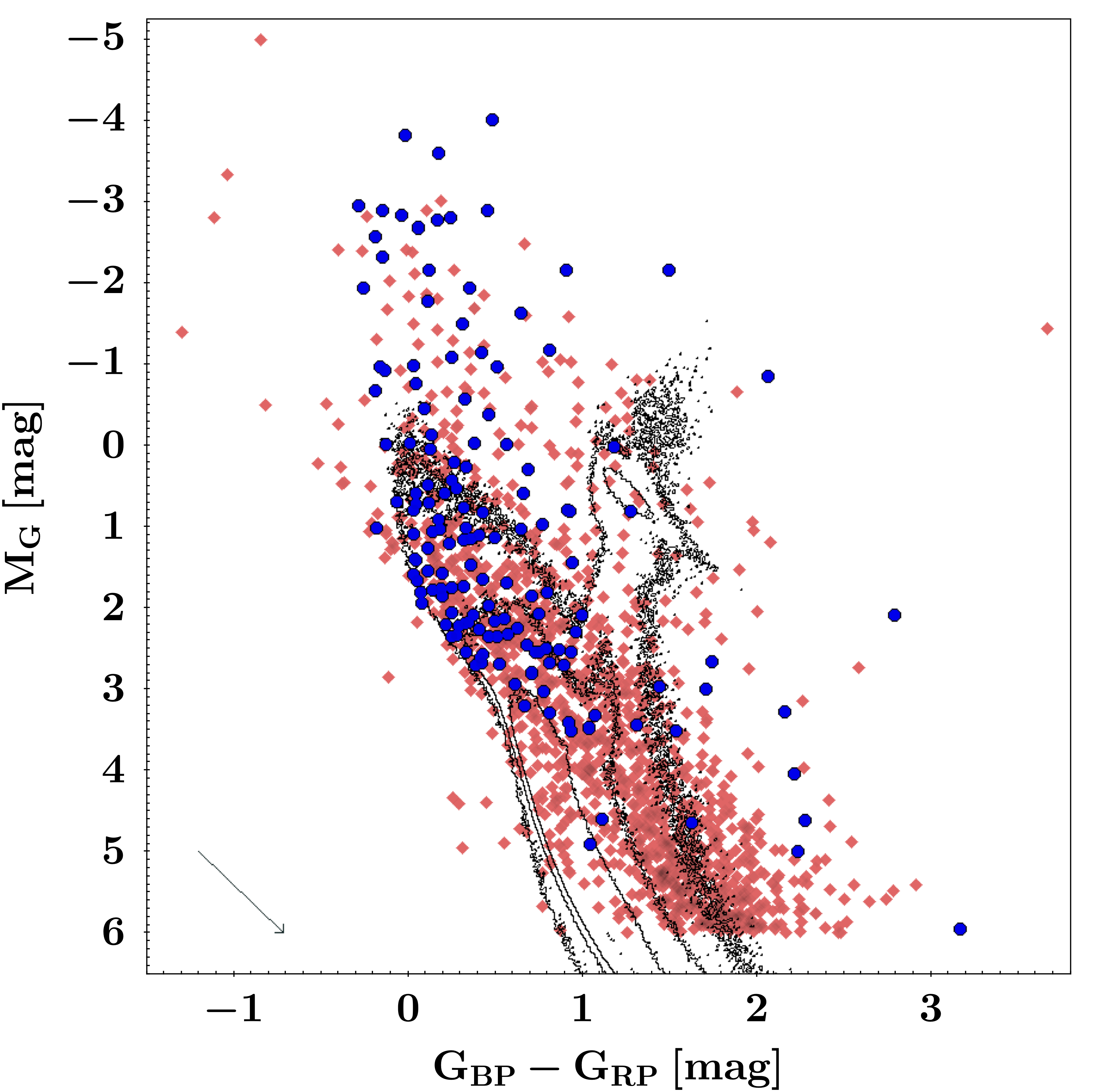}
\caption{\textit{Gaia} colour vs. absolute magnitude HR diagram. Blue dots are previously known Herbig Ae/Be stars with good astrometric solution corrected from interstellar extinction. Red diamonds are new Herbig Ae/Be candidates corrected from interstellar extinction. An extinction vector corresponding to $A_{G}=1$ is shown on the bottom left. Black contours trace \textit{Gaia} sources within 500 pc with good astrometric solution.}\label{Progress}
\end{figure}

We have constructed a new catalogue of PMS candidates, of which some can be plotted accurately in the HR diagram (Fig. \ref{HR_grid} at top right), a selection-independent plot. Therefore, we can further select the Herbig Ae/Be stars. In order to do this we study where the known HAeBes (Sect. \ref{T_Herbig}) are placed in the HR diagram using the same quality constrains and extinction corrections. This is done in Fig. \ref{Progress}. There, we estimate that PMS candidates with absolute magnitude $M_{G}<6$ are possible HAeBe candidates. This is taking into account that the intrinsic extinction, typically large for these objects, has not been considered (most of these sources do not have measured spectral types). This way, we retrieve 1361 new Herbig Ae/Be candidates which are marked in Table \ref{Table2} (end of the pipeline, Fig. \ref{fig:Pipeline2}). This constitutes an improvement of one order of magnitude with respect to the $\sim273$ previously known HAeBes (\citealp{2018A&A...620A.128V}). The new Herbig Ae/Be candidates are shown in Fig. \ref{Progress}. 

By construction, these HAeBe candidates are astrometrically well behaved sources (RUWE<1.4 and $\varpi/\sigma(\varpi)\geq10$). Hence, we expect many more HAeBes among the PMS candidates, as most of them do not satisfy these conditions ($\sim 60\%$, see Fig. \ref{HR_grid} at top left) or do not even have parallax information ($\sim 6\%$). For example, using a more relaxed $\varpi/\sigma(\varpi)\geq5$ parallax constraint gives 2226 HAeBe candidates, but their distance and interstellar extinction uncertainties do not allow us to separate them as nicely from the low-mass candidates. In contrast, using less precise parallaxes allow to retrieve candidates at farther distances and typically more massive. As the list of all PMS candidates is available in Table \ref{Table2}, future studies may want to use less conservative thresholds to the astrometric quality and select their own set of Herbig Ae/Be candidates.

\subsection{Variable candidates}\label{S_Variability}
UX Ori type objects (UXORs) are sources with irregular brightness variations from 2-3 magnitudes in the optical. Observed light gets bluer in the deep minima, and the fraction of polarised light increases. Many of them are catalogued as HAeBes and their extreme variability is explained by eclipsing dust clouds in nearly edge-on sources and scattering radiation in the circumstellar environment (\citealp{1997ApJ...491..885N}; \citealp{2000ASPC..219..216G}; \citealp{2000A&A...364..633N}; \citealp{2001A&A...379..564O}). In \citet{2018A&A...620A.128V} we provided strong support to the edge-on disc explanation using H$\alpha$ line profiles of known HAeBes. In addition, using a similar variability indicator to $G_{var}$ (Eq. \ref{Gvar}), we found that all catalogued UXORs with V band detected variabilities above 0.5 mag are strongly variable (17 objects). This implies that these indicators effectively trace irregular photometric variability.

By using a variability threshold, in \citet{2018A&A...620A.128V} we proposed 31 new UX Ori candidates among the previously known HAeBes. The equivalent $G_{var}$ threshold is $G_{var}\geq10$. PMS candidates with $G_{var}\geq10$ are marked in Table \ref{Table2} with the flag of `Var' (3436 sources) and for the HAeBe candidates the UXOR phenomenon is the most likely explanation. As we are tracing variability by using the \textit{Gaia} DR2 uncertainties, sources without intrinsically irregular photometry like binaries or extended objects can pop out as strongly variable.

This means that $\sim41\%$ of the new PMS candidates are of this variability type. This proportion increases to $\sim49\%$ when it comes to the HAeBe candidates. This number is consistent with the UXOR behaviour caused by an inclination effect (50\% predicted by \citealp{2000A&A...364..633N}). Very probable PMS candidates are in general very variable (see Sects. \ref{Test_eval} and \ref{S_Evaluation}), so most of the best PMS candidates appear as `Var' in Table \ref{Table2}.

Another assessment of our variability proxies can be achieved by cross-matching the PMS candidates with variability surveys. A 5 arcsecond cross-match with ASAS-SN (\citealp{2019MNRAS.485..961J}) gives 949 sources, of which 830 (87\%) have $G_{var}\geq10$. In addition, 557/949 (59\%) appear as of Young-Stellar Object (YSO) variability type. A 5 arcsecond cross-match with the Zwicky Transient Facility (ZTF, \citealp{2019PASP..131a8003M}) gives 6438 sources, including 95\% of those with $G_{var}\geq10$ in the sky region covered by the survey. A 5 arcsecond cross-match with ATLAS-VAR (\citealp{2018AJ....156..241H}) gives 2216 objects. Of these, 1960 (88\%) have variabilities which are hard to classify by the machine classifier, suggesting that they are likely of an irregular type, similar to those of PMS sources. Finally, if we cross-match our results with the catalogue of long-period variable candidates of \citet{2018A&A...618A..58M}, which also contains a small set of YSO candidates, we obtain 491 matches with the set of PMS candidates, of which 444 (90\%) have $G_{var}\geq10$. According to \citet{2018A&A...618A..58M} classification, 297/491 (60\%) are YSO candidates but 190/491 (39\%) are long-period variable candidates (4/491 are undetermined because they lack parallaxes). These possible contaminants are addressed in Sect. \ref{S_Evaluation}.

\subsection{Comparison with \citet{2019MNRAS.487.2522M} and other catalogues and surveys}\label{S_Marton}

\citet{2019MNRAS.487.2522M} did a similar study to the one presented here but looking for YSO in general and only using \textit{Gaia} DR2, 2MASS, WISE colours and passbands, and the optical depth from the \textit{Planck} dust opacity map. Therefore, they did not use H$\alpha$ or variability information and did use distance-dependent features. In addition, they restricted the search to high dust opacity regions. They found 1,768,628 potential new YSO candidates (with the recommended $p\geq0.9$) using a Random Forest algorithm. Giving the differences between the two approaches in terms of considered sources, training data, and features; it should not be surprising if there are not many objects in common between the two studies and yet both are highly accurate. However, we find that $48\%$ of our PMS candidates are within the \citet{2019MNRAS.487.2522M} catalogue. Moreover, this percentage slightly increases at higher probability thresholds of our catalogue ($56\%$ at $p\geq0.95$). Regarding the Herbig Ae/Be candidates (see Sect. \ref{Herbig_candidate}), $56\%$ of them are present among the YSOs of \citet{2019MNRAS.487.2522M}. In contrast, only $11\%$ of our catalogue of classical Be stars appear as YSO in \citet{2019MNRAS.487.2522M}. When moving to $p\geq0.85$ this number goes down to $0\%$.

This is a good assessment of the quality of our categorisation as an independent study, using a different algorithm and training data, has achieved relatively similar results regarding PMS sources (taking into account the differences between methodologies) but has not found almost any of our CBe candidates (and none of the best CBe candidates). This, in addition to support our selection, proves that our HAeBe candidates are nicely separated from the population of CBe stars. The differences between the two studies probably lie in that we are using H$\alpha$ and variability information and that \citet{2019MNRAS.487.2522M} searched only in dusty environments, being this way position-dependent. In addition, we demand detections up to W4 ($22 \mu m$), whereas these authors only demand detections up to W2 ($4.6 \mu m$). Further assessment is that, as in \citet{2019MNRAS.487.2522M}, we find that 62 of our PMS candidates are within the \textit{Gaia} Photometric Science Alerts published at the time of writing (a project that looks for transient events in the \textit{Gaia} data, \citealp{2019ASPC..523..261D}); 13 of them appear as YSO, 47 as unknown, and only two appear as non-PMS. Conversely, of the 87 YSO in the Alerts, 18 are in the SoSt, which means that we only missed five that were classified as `other source'.

Similarly, of the PMS candidates in SIMBAD (2607 within 1 arcsecond cross-match at the time of writing) 974 ($\sim 37\%$) appear catalogued as PMS or PMS candidate. There are 18 objects appearing as CBe, but these were mostly catalogued by \citet{2008MNRAS.388.1879M} and \citet{2016A&A...591A.140G}. These papers selected CBes using simple cuts in IPHAS+, 2MASS, or WISE observables and hence we understand that our analysis supersedes theirs. In addition to this, 663 sources ($\sim 25\%$) appear as with emission lines, infrared bright, or variable. Only 356 sources ($\sim 14\%$) appear as clearly non-PMS. This includes 101 AGB candidates and 16 carbon star candidates that are addressed in Sect. \ref{S_Evaluation}. As explained in that section, we expect this number of 356 PMS candidates classified as non-PMS by other studies to be considerably lower, so this cross-match with SIMBAD is consistent with the estimated precision in Sect. \ref{S_out} of $P\gtrsim80\%$ for the catalogue of PMS candidates. The other 596/2607 sources do not have a defined category in SIMBAD. \object{VES 263}, the new Herbig Ae/Be star discovered by \citet{2019MNRAS.488.5536M} is not within the SoSt.

Of the classical Be candidates in SIMBAD (280 within 1 arcsecond cross-match at the time of writing) 17 appear as CBe (again, most from \citealp{2008MNRAS.388.1879M} and \citealp{2016A&A...591A.140G}) and 197 as with emission lines. Only nine are clearly not CBes, of which four are of PMS nature and three appear as variable. This reinforces the idea that the algorithm is efficiently separating PMS sources from classical Be stars. The other 57/280 sources do not have a defined category in SIMBAD.

Finally, using a cross-match aperture of 20 arcsecond we find 26 matches between the set of PMS candidates and the Gaia-ESO Public Spectroscopic Survey (\citealp{2012Msngr.147...25G}). A fraction of 24/26 sources have hydrogen lines in emission: 14/26 show double-peaked emission (although two might be considered P-Cygni), 6/26 single-peaked emission, 3/26 are either single-peaked or double-peaked, and one shows a clear inverse P-Cygni profile. Only 2/26 spectra have H$\alpha$ line in absorption.  The line profile fractions agree with those studied in \citet{2018A&A...620A.128V} for known HAeBes ($31\%$ single-peaked, $52\%$ double-peaked, and $17\%$ P-Cygni). This gives independent spectroscopic evidence for the PMS nature of the new PMS candidates.

\section{Quality assessment}\label{Quality assessment}

Table \ref{Table_summary} summarises the final number of sources in the resulting catalogues of PMS and CBe candidates. Table \ref{Table_summary} also indicates the number of known sources considered in Sect. \ref{Trainingset}, of which those having all observables were used for the set of Labelled Sources. In this section we evaluate the classification from different perspectives and give insights on the relative importance of the different observables used for the selection. In addition, we discuss any detected bias or flaw in the final catalogues of PMS and classical Be candidates. In general, these mostly affect sources with a bad astrometric solution in \textit{Gaia} DR2 so they do not implicate the catalogue of new Herbig Ae/Be candidates (Sect. \ref{Herbig_candidate}).

\subsection{Classification on the test sets}\label{Test_eval}
One way to analyse the classification is to study the evaluation on the test sets. As described in Sect. \ref{Bootstrap}, we evaluated the performance of the ANN in 30 different test sets. As the selection of the test set is random in every iteration, almost all of the known PMS and classical Be sources were in the test set at some point. If we average these 30 evaluations we end up with 793 PMS and 733 CBe known sources that have been independently assessed by the algorithm.

Regarding the classification of known PMS sources, the most noteworthy trend is that very variable PMS stars in either indicator ($G_{var}$ and $V_{htg}$) are identified. Although those known PMS stars with $r-H\alpha>1.3$ are identified, objects with $0<r-H\alpha<1.3$ are spread over the whole range of probabilities. Thus, $r-H\alpha$ does not seem to play an important role in detecting PMS sources (see Sect. \ref{Important Features}). This also happens with $G_{BP}-G_{RP}$. However, in these two cases, known PMS sources with low $r-H\alpha$ or bluer are those who tend to be given high CBe probabilities. The known PMS sources that were not identified were mostly stars with low near-IR excess ($J-K_{s}$), which are also the ones that are given high CBe probabilities. This is probably because these are more similar to CBe stars. Surprisingly, we miss many known PMS sources with high mid-IR excess ($W1-W4$) and those that had very low $W1-W4$ values were mostly not identified, which again are the ones with higher CBe probabilities. In general, very few known PMS sources are assigned to the CBe category, although many known PMS sources are not classified as such (algorithm recall on the PMS group is $R=78.8\pm1.4\%$, Sect. \ref{Evaluation on HR}). 

Regarding the known classical Be sources, the algorithm also identifies the very variable ones as CBe. This implies that it uses variability to differentiate PMS and CBe sources from other objects. CBe sources with high $r-H\alpha$ are normally given high PMS probabilities but in general they are not misclassified. There is no trend between $r-H\alpha$ values and CBe assigned probabilities. In contrast, there is a trend with $G_{BP}-G_{RP}$ and redder objects are less likely to be classified as CBe and are given higher PMS probabilities, but are rarely misclassified as PMS. In addition, CBe sources show no CBe probability trend with $J-K_{s}$ or $W1-W4$ although sources with $W1-W4\gtrsim3$ are normally not classified as CBe. Similarly, CBe sources with higher near- and mid-IR excesses are given higher PMS probabilities but are infrequently assigned to the PMS category.

Evaluation on the test sets indicates that the algorithm effectively identifies sources of different categories and uses the various observables to trace the main characteristics of PMS and classical Be stars.

\begin{table}
\caption{Summary of the number of known sources of each type considered together with those included in the set of Labelled Sources because of having all observables. The last column indicates the number of new candidates of each type classified by the algorithm.}
\label{Table_summary}      
\centering                                      
\begin{tabular}{l c c c}          
\hline\hline
\addlinespace[0.07cm]
 & Considered & Labelled Sources & Classified\\ 
& sources & set  & $p\geq0.5$\\
\hline 
   \addlinespace[0.07cm]
   Herbig Ae/Be & 255 & 163 & - \\
   T-Tauri & 3171 & 685 & - \\
   PMS & 3426 & 848 & 8470\\
   Classical Be & 1992 & 775 & 693\\
\hline      
\end{tabular}
\tablefoot{To be considered, we demanded the sources to be present in \textit{Gaia} DR2. For the algorithm there is just a single class of PMS objects, which is constructed by combining known Herbig Ae/Be and T-Tauri stars. Herbig Ae/Be and T-Tauri candidates can be selected from the set of PMS candidates using the HR diagram. In Sect. \ref{Herbig_candidate} 1361 Herbig Ae/Be candidates were obtained this way.}
\end{table}

\subsection{Final catalogues assessment}\label{S_Evaluation}

In the following points we discuss a few biases and flaws detected in the final catalogues of PMS and CBe candidates:

\begin{enumerate}

\item We demand to have detections up to W4 ($22 \mu m$) and in the H$\alpha$ passbands. Although we are training with sources that span the whole range of values in these observables, this induces some biases as we are excluding in the training many of the less extreme sources and hence biasing the posterior selection. This is aggravated given that the H$\alpha$ EW to $r-H\alpha$ colour conversion of Sect. \ref{Trainingset} can only be applied to sources with observed emission above the continuum. This effect can be quantified if we compare the output catalogues with all the known sources gathered in Sect. \ref{Trainingset} (see Table \ref{Table_summary}). This way, the mean value of $r-H\alpha$ for known PMS (classical Be) sources (using one standard deviation as error) is $r-H\alpha=0.74\pm0.36$ mag ($0.38\pm0.18$ mag) and for the candidates is $r-H\alpha=0.87\pm0.46$ mag ($0.63\pm0.20$ mag). The mean value of $W1-W4$ for known PMS (classical Be) sources is $W1-W4=4.0\pm2.2$ mag ($1.7\pm1.3$ mag) and for the candidates is $W1-W4=5.2\pm1.4$ mag ($2.24\pm0.71$ mag). In short, the retrieved candidates are the more extreme of their kind in terms of H$\alpha$ emission and IR excess (specially mid-IR excess). This particularly affects the catalogue of CBe candidates, as these have typically less extreme values. In Fig. \ref{Massive_grid} we present the frequency density distribution of the final catalogues of PMS and CBe candidates for a subsection of key observables ($G_{BP}-G_{RP}$, $J-K_{s}$, $W1-W4$, $r-H\alpha$, $G_{var}$, and $V_{htg}$) together with the distribution of all known sources.

\item As mentioned in Sect. \ref{S_Data}, WISE presents many spurious photometric detections in the Galactic plane. To investigate this, \citet{2014ApJ...791..131K} used a set of AllWISE quality parameters and additional selection criteria to determine that only $\sim 28\%$ of the sources in their study have reliable W3 and W4 detections. \citet{2019MNRAS.487.2522M}, using a different approach, concluded that only 10\% of their set have reliable W3 and W4 photometry. These authors used very stringent criteria for the sake of purity and these percentages may be slightly pessimistic.

We decided to use these passbands because of their expected importance in separating CBes from PMS sources (see Sect. \ref{Important Features}). A more relaxed constraint, using the extended source flag of AllWISE distinct to 0 gives 44\% and 27\% of badly behaved PMS and CBe candidates respectively (marked in Tables \ref{Table2} and \ref{Table3} with a `W3W4' flag). We note that, in contrast to \citet{2019MNRAS.487.2522M}, we are using many observables in addition to W3 and W4 so the algorithm can deal better with these being spurious.

\begin{figure*}[ht!]
\centering\includegraphics[scale=0.741]{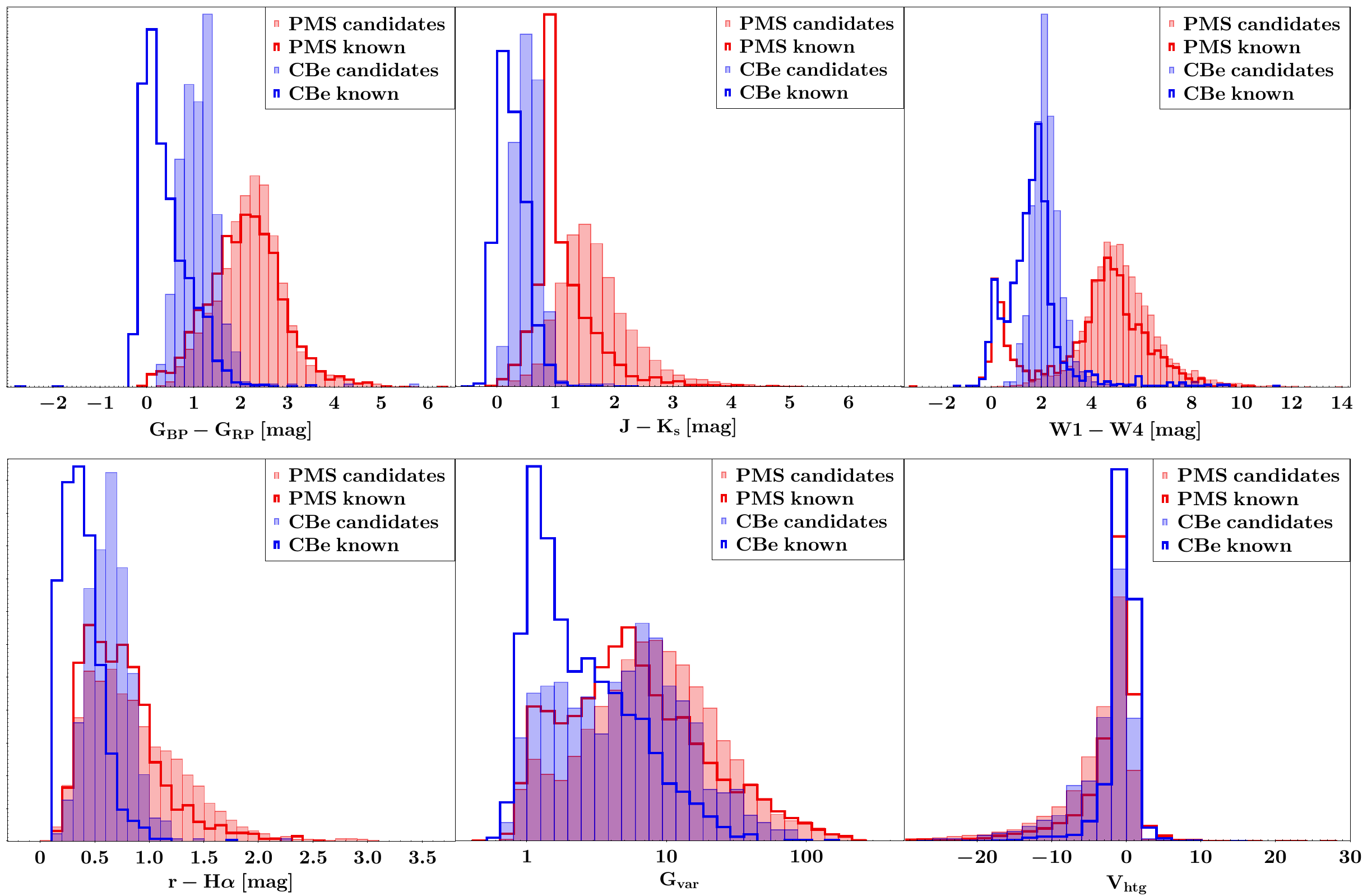}
\caption{Frequency density distribution of PMS candidates in shaded red and classical Be candidates in shaded blue for different selected observables. The red and blue lines respectively trace all considered known PMS and CBe objects, including those without all the observables. Area of each histogram has been normalised to one. For clarity, some individual extreme sources are out of bounds in the $r-H\alpha$, $G_{var}$, and $V_{htg}$ plots.}\label{Massive_grid}
\end{figure*}

\item As described in Sect. \ref{S_Data}, because of the cross-match, we estimated that 1/42 (1/625) sources of the SoSt on average are fake, in the sense that their associated IR ($H\alpha$) photometry do not belong to them. However, only 17 (6) PMS and no CBe candidates appear with duplicated IR ($H\alpha$) information. The sources that have the AllWISE, or IPHAS or VPHAS+ name repeated in the SoSt are marked in Tables \ref{Table2} and \ref{Table3} with the `ID AllW' or `ID IPH/VPH' flag respectively.

\item There are 104 SIMBAD AGB stars in the catalogue of PMS candidates (only three appear as of confirmed AGB nature and the rest appear as candidates). This is because they were all classified in one single paper (\citealp{2008AJ....136.2413R}), where they attempted to separate YSO from AGB stars using a simple colour and magnitude selection criteria in the near- and mid-IR. We understand that these are contaminants in that work as our analysis supersedes theirs.

\item We have detected a high incidence of Planetary Nebulae (PN) detected as PMS candidates. Observational similarities between YSO, B[e] stars, and PN have been reported before (e.g. \citealp{2010PASA...27..129F}; \citealp{2012IAUS..283..316B}; \citealp{2015EAS....71..181M} and references therein). This is mainly caused because PN show high $r-H\alpha$ colours. In addition, as they are extended they present large \textit{Gaia} uncertainties, so they can appear as highly photometrically variable in our indicators (Eqs. \ref{Gvar} and \ref{Var_col}). Of the PMS candidates in SIMBAD (Sect. \ref{S_Marton}), there are 57 ($\sim 3.5\%$) catalogued as PN and 34 as possible PN. By studying their location in the observable space we concluded that any candidate with a $r-H\alpha\gtrsim1.3$ should be treated with caution ($16\%$ of the sample of PMS candidates). Below that number we estimate the possible contamination by PN to be below $5\%$. Candidates with $r-H\alpha\geq1.3$ are marked in Tables \ref{Table2} and \ref{Table3} with a `PN' warning flag. Indeed, there are eight PN (within 5 arcsecond) from \citet{2014MNRAS.443.3388S}, 40 from \citet{2003A&A...408.1029K}, and three from \citet{2008ApJ...689..194S} among our PMS candidates. A fraction of 46/51 (90\%) have $r-H\alpha\geq1.3$. We expect also some contamination in these works. PMS candidates with $r-H\alpha\geq1.3$ have mostly absolute magnitudes $M_{G}>6$, so they barely contaminate the sample of Herbig Ae/Be candidates (Sect. \ref{Herbig_candidate}). In contrast, the few candidates with a `PN' warning flag and $M_{G}<6$ have a significant probability of being unclassified B[e] (FS CMa) stars.

\item Similarly, we detect a high number of carbon stars among our PMS candidates (71 confirmed and 16 candidates, according to SIMBAD). They stand out in variability and near-IR excess, but not in mid-IR excess, where they have a smaller excess than the rest of the candidates. Only two were classified as Herbig Ae/Be (\object{Gaia DR2 1828276425855506304} and \object{Gaia DR2 5336019093122634624} with PMS probability of 0.52 and 0.53 respectively) as the other 85 do not have reliable astrometry. Not surprisingly, 80/87 (92\%) were identified as contaminants in Appx. \ref{S_Visualization} and marked in Table \ref{Table2} with a `G\textsubscript{UMAP}' warning flag (see Appx. \ref{S_Visualization} for further details). Therefore, we do not expect them to have a high impact on the final catalogue of PMS candidates. In addition, 29 PMS candidates appear as variable stars of Mira Cet type in the cross-match with SIMBAD. A fraction of 17/29 (59\%) are flagged as `G\textsubscript{UMAP}' in Table \ref{Table2}.

\item 51 PMS candidates are in the catalogue of OH/IR stars of \citet[within 5 arcsecond]{2015A&A...582A..68E}. These 51 sources have different categories in SIMBAD and may have been assessed in previous points and sections as of other nature (even YSO). From the catalogue of AGBs of \citet{2017JKAS...50..131S}, 26 (within 5 arcsecond) are among our PMS candidates (14, 54\%, are flagged as `G\textsubscript{UMAP}' in Appx. \ref{S_Visualization}). Similarly, a very small fraction of the CBe catalogue is potentially contaminated by sources with very strong mid-IR excess or that seem evolved. There are 0 post-AGB stars of \citet{2007A&A...469..799S} in either catalogue.

\end{enumerate}

Most of the contaminants discussed in the previous points can be avoided by constraining the position in the HR diagram and moving away from the giant and supergiant region. One way to do so is with the constraint applied in Fig. \ref{HR_grid} at right. This also implies that the sample of Herbig Ae/Be candidates (see Sect. \ref{Herbig_candidate}) is barely affected by these contaminants. Conversely, the HR diagram is a powerful tool to discard contaminants in other catalogues which were correctly classified in this work.

\subsection{Probability of being either PMS or classical Be}\label{Sect_to_either}

There are 1309 sources whose probabilities of being PMS and CBe sum up to a probability $\geq50\%$, but that are below $50\%$ in either category (i.e. $p(PMS)+p(CBe)\geq50\%$ but $p(PMS)<50\%$, $p(CBe)<50\%$, see Fig. \ref{fig:Probability_map}). This means that the algorithm thinks that they belong to one of those two categories, but it is unable to say which. A closer look at these objects reveals that they behave very similarly to known CBe stars in terms of $G_{var}$, $V_{htg}$, and $r-H\alpha$ but their mean $G_{BP}-G_{RP}$ colour is redder and they have slightly larger near-IR excess ($J-K_{s}$). The $W1-W4$ colour peaks where CBe stars do but it is quite homogeneously distributed. This is not surprising as in Fig. \ref{projection} at left they appear mostly mixed with the CBe candidates, in the region where the PMS candidates that are more similar to known CBes are placed.

These borderline objects are interesting in their own right and contain probably most of the less active PMS sources and in particular, most of the less active and probably more evolved Herbig Ae/Be stars. These sources are listed in a table equivalent to Tables \ref{Table2} and \ref{Table3} which is only available in electronic form.

\begin{table*}
\caption{Evaluation of the impact of the different observables in the final selection. }
\label{Table_less_observables}      
\centering                                      
\begin{tabular}{l c c c c c c c c c c}          
\hline\hline
\addlinespace[0.07cm]
& Precision & Recall & Precision & Recall & PMS & CBe & PMS & CBe\\ 
& PMS (\%) & PMS (\%) & CBe (\%) & CBe (\%) & ($p \geq0.5$) & ($p \geq0.5$) & P/R (\%) & P/R (\%)\\
\hline
   \addlinespace[0.07cm]
   \textbf{All observables} & \pmb{$40.7\pm1.5$} & \pmb{$78.8\pm1.4$} & \pmb{$88.6\pm1.1$} & \pmb{$85.5\pm1.2$} & \pmb{$8470$} & \pmb{$693$} & \pmb{$100/100$} & \pmb{$100/100$}\\
   No $r$, $H\alpha$ & $35.5\pm1.4$ & $73.1\pm1.5$ & $87.1\pm1.1$ & $85.52\pm0.97$ & 9108 & 718 & 77/82 & 80/83\\
   No $G_{BP}$, $G$, $G_{RP}$ & $32.11\pm0.99$ & $74.2\pm1.2$ & $87.5\pm1.0$ & $82.2\pm1.3$ & 10772 & 650 & 63/80 & 80/75\\
   No $J$, $H$, $K_{s}$ & $32.4\pm1.2$ & $81.9\pm1.4$ & $87.42\pm0.84$ & $86.59\pm0.64$ & 11123 & 758 & 69/91 & 81/88\\
   No $W1$, $W2$, $W3$, $W4$ & $28.33\pm0.68$ & $80.4\pm1.3$ & $70.2\pm1.5$ & $76.6\pm1.5$ & 13303 & 1890 & 47/74 & 25/69\\
   No $W1$, $W2$ & $37.6\pm1.1$ & $75.1\pm1.3$ & $86.0\pm1.2$ & $85.8\pm1.1$  & 8398 & 742 & 76/75 & 75/80\\       
   No $W3$, $W4$ & $31.7\pm1.1$ & $81.0\pm1.2$ & $83.9\pm1.2$ & $75.4\pm1.9$ & 11889 & 820 & 68/95 & 58/69\\ 
   No $G_{var}$, $V_{htg}$ & $23.49\pm0.66$ & $74.0\pm1.3$ & $91.98\pm0.83$ & $80.7\pm1.1$ & 18055 & 358 & 40/84 & 100/52\\
   No $G_{var}$ & $27.78\pm0.84$ & $74.7\pm1.4$ & $92.62\pm0.80$ & $80.2\pm1.8$ & 13667 & 355 & 53/85 & 99/51\\
   No $V_{htg}$ & $40.0\pm1.1$ & $73.1\pm1.4$ & $91.89\pm0.97$ & $83.8\pm1.6$ & 7543 & 468 & 97/87 & 100/67\\
\hline      
\end{tabular}
\tablefoot{Metrics and number of sources obtained for the different classifications excluding the indicated observables. Precisions are lower limits (see Appx. \ref{Weights}). Last two columns indicate the `precision' and `recall' of these new catalogues with respect to the catalogues obtained when using all the observables. This is, the proportion of sources of the different catalogues that are in the catalogues obtained using all the observables and the proportion of the candidates obtained when using all the observables that are still retrieved when excluding the respective observables.}
\end{table*}

\subsection{Important observables}\label{Important Features}

In this section, we try to assess how important the different observables are for identifying PMS and classical Be stars. This is not trivial because of the intrinsic nature of the ANN-based algorithm, as the selection itself is not an obvious process. 

What we did was to repeat the pipeline explained in Appx. \ref{S_Description of the pipeline} excluding some observables. We did not include the sources that were removed by demanding detections in those observables (see Sect.  \ref{Trainingset}), as this would make it impossible to know whether the new results are caused by the new sources or the lack of those observables. Similarly, the ANN architecture was optimized for the whole set of observables (Appx. \ref{architecture}), so using fewer observables might alter the performance in an uncontrolled manner that can affect our conclusions. In order to minimise the impact of this we removed only a few observables at a time. Another problem is that by using fewer observables without changing the complexity of the algorithm we may start overfitting the selection. To assess this we checked that the retrieved precisions and recalls are within reasonable limits in each case. 

In Table \ref{Table_less_observables} we present the results (precision, recall, PMS and CBe candidates with $p\geq0.5$) obtained when applying the same algorithm of Sect. \ref{S_Description of the pipeline} to the same sources of Sects. \ref{S_Data} and \ref{Trainingset} but excluding certain observables (in the case of passbands this implies excluding the colours they appear on, see Sect. \ref{Observables}). In the last two columns we express the percentage of PMS and CBe candidates of those selections that were also retrieved when using all the observables and the percentage of sources classified when using all the observables present in these new catalogues. These two values, in some sense, are equivalent to precisions and recalls if we assume the catalogues obtained using all observables as reference. As the algorithm was optimized to maximise the precision on PMS sources (see Appx. \ref{Traing_Test}), this is maximum when using all observables. Similarly, when applying the algorithm to a smaller set of observables the results are going to be inevitably worse (as we do not include more sources). However, there is information in how much worse they get, although we can only talk in relative terms.

As outlined above, this table should be treated with caution but it gives information about the relative importance of the different observables in the selection of PMS and CBe candidates. We discuss the main outcomes of Table \ref{Table_less_observables} in the following points:

\begin{enumerate}

\item Not using $r-H\alpha$ does not change the output tremendously. The number of PMS and CBe candidates retrieved is similar and there is only a small discrepancy (of $\sim20\%$) with the case of using this colour, in the sense that mostly the same sources are identified and not many sources that were not identified when using $r-H\alpha$ are included. This is because cool stars have the same $r-H\alpha$ colour than hot stars with emission, see \citet{2005MNRAS.362..753D}, and hence this observable is not efficient in separating PMS and CBe sources from other objects.

\item $G_{BP}$, $G$, and $G_{RP}$ are more relevant for the selection. If we do not use them the discrepancy with the original set of PMS candidates obtained using all the observables is higher than in the $r-H\alpha$ case. Many more sources are obtained (which we can consider a sort of contaminants) without losing many of the catalogued ones using all observables. In the case of the CBes, the effect is the opposite, not many contaminants are added but a few of the sources identified with these colours are lost. This might be caused by these colours carrying the photospheric information less affected by disc emission and hence more representative of temperature. Therefore, for the PMS case including them helps to remove candidates with unfeasible temperatures (like white dwarfs) and in the CBe case it helps a bit the selection as they are mostly blue with low extinctions (see Figs. \ref{fig:Train} and \ref{HR_grid}). 

\item $J$, $H$, and $K_{s}$: Not using these 2MASS passbands makes us lose very few PMS and CBe candidates (less than in the previous cases) but we get many PMS contaminants, implying that the colours involving these passbands are relevant for differentiating PMS sources from other objects, although they are not critical for the selection. These observables do not seem to have a big impact for the classification of CBes.

\item As expected, $W1$, $W2$, $W3$, and $W4$ are very important. Not using these WISE passbands drastically increases the number of contaminants and significantly reduces the number of PMS and CBe candidates obtained when compared to the case of using this information. However, removing so many observables at a single time can cause the algorithm to start overfitting, so these results might be a bit exaggerated. If we choose to not use only $W1$ and $W2$ the selection is not much affected (only a bit more than in the $r-H\alpha$ case). Not surprisingly, it is if we choose not to use $W3$ or $W4$ when we obtain very poor results. We retrieve mostly the same PMS candidates but also many PMS contaminants. The larger impact is in the case of the CBes, as we miss a lot of them and misclassify almost half of the obtained catalogue. This is expected, as at this wavelength range is where the discs of PMS stars and CBes start differing. Therefore, probably many CBes are moved to the PMS catalogue lowering its precision. This is the reason we opted to keep these passbands even though they suffer from a high incidence of spurious detections (see Sects. \ref{S_Data} and \ref{S_Evaluation}).

\item $G_{var}$ and $V_{htg}$: The observable $G_{var}$ proves to be of the utmost importance for identifying PMS sources. When excluding both variability indicators, we get twice as many PMS candidates than in the catalogue using all observables, with an almost full recovery of the later ones. Curiously, half of the CBes are lost, but not a single contaminant is added. If we exclude them independently we find that $V_{htg}$ is almost irrelevant. It only helped to classify several CBes. In contrast, not using $G_{var}$ doubles the number of PMS candidates retrieved (so half the sources can be considered contaminants) and halves the number of CBes obtained. The number of CBe contaminants is close to zero in every case. All these imply that this indicator is very useful for separating PMS sources from other objects and, in some cases, to differentiate them from CBes, but ineffective to identify CBes from the background sources. This is just as expected as we know from \citet{2018A&A...620A.128V} that $G_{var}$ mostly traces irregular photometric variations caused by edge-on dusty discs. 

\end{enumerate}

It is clear that if we had optimized the algorithm for each situation using all the sources available in each case for the training we would have obtained more candidates and, from Table \ref{Table_less_observables}, it is safe to say that these would have been more contaminated.

\section{Conclusions}\label{S_conclusions}
In this work we have used Machine Learning techniques (mainly artificial neural networks) to produce a catalogue of new PMS candidates and a catalogue of new classical Be stars from 4,150,983 sources resulting from the cross-match of \textit{Gaia} DR2, AllWISE, IPHAS, and VPHAS+. To each of the 4,150,983 sources we assigned a PMS and a classical Be probability. The entire set of sources is available in electronic form so the users can choose the probability thresholds ($p$) that fit their needs. The categorisation is distance and position independent. We have given independent evidence that the categorisation is accurate and consistent, having a high efficiency at separating PMS sources from classical Be stars.

\begin{enumerate}
\item  At $p\geq50\%$ the catalogue of PMS candidates is: 8,470 sources, recall (completeness) of $78.8\pm1.4\%$ and lower limit to precision of $40.7\pm1.5\%$. Independent analyses indicate that the real precision is around double this value. The PMS candidates are distributed all over the Galactic plane, tend to be associated with nebulosities and appear mostly in PMS locations in the HR diagram. This catalogue (Table \ref{Table2}) is available in electronic form independently.

\item Out of the PMS candidates, 2052 have a good astrometric solution in \textit{Gaia} DR2 (RUWE<1.4 and $\varpi/\sigma(\varpi)\geq10$). Of those, 1361 have a location in the HR diagram compatible with that of known Herbig Ae/Be stars. Many more Herbig Ae/Be candidates can be obtained from the set of PMS candidates by relaxing the constraint to the parallax quality. This comes at a price, as the larger errors on the absolute magnitudes make it more difficult to distinguish low and high mass objects from each other.

\item At $p\geq50\%$ the catalogue of classical Be candidates is: 693 sources, recall (completeness) of $85.5\pm1.2\%$ and lower limit to precision of $88.6\pm1.1\%$. The classical Be candidates are distributed all over the Galactic plane and appear mostly in classical Be locations in the HR diagram.  This catalogue (Table \ref{Table3}) is available in electronic form independently.

\item There are 1309 sources that have a combined probability of larger than $50\%$ of belonging to either of these categories but each individual category has a probability of below $50\%$. In general these objects have characteristics of classical Be stars or the less extreme PMS sources in the observables used. These sources are listed in a table equivalent to Tables \ref{Table2} and \ref{Table3} which is only available in electronic form.

\item We have made a thorough analysis of the possible biases and contaminants present in the selection. The biases can be summarised in that we are retrieving the most extreme PMS and classical Be sources in the observables used. The contaminants are mostly giants, with the special case of Planetary Nebulae appearing as PMS. These contaminants are sparse and easy to avoid. Instructions are given to minimise their impact (in Sects. \ref{S_out}, \ref{Test_eval}, \ref{S_Evaluation}, and Appx. \ref{S_Visualization}). The new HAeBe candidates are very little affected by these contaminants, mainly as by construction they have a good astrometric solution.

\item 3436 PMS candidates (at $p\geq50\%$) show strong irregular photometric variabilities. For the HAeBe candidates the UXOR phenomenon is the most likely explanation. The proportion of variable HAeBe candidates is consistent with the inclination explanation for the UX-Ori type variability. 

\item An analysis of the relative importance of the different observables used shows that irregular photometric variability is extremely important for identifying PMS sources among other objects and W3 $[12 \mu m]$ and W4 $[22 \mu m]$ are very powerful to separate PMS sources from classical Be stars. On the other hand, $r-H\alpha$ is not very relevant for the selection of these two types of objects.

\end{enumerate}

These new catalogues of PMS and classical Be candidates will be subjected to follow up studies of their properties using independent spectroscopic observations. The catalogue of new PMS candidates was accepted as target list for the WEAVE survey, a wide-field spectroscopic survey which will be carried out at the William Herschel Telescope in the forthcoming years (\citealp{2018SPIE10702E..1BD}).

\begin{acknowledgements}
We thank A. S. Miroshnichenko for his help with the list of known FS CMa stars. This project benefited from discussions in the Gaia DR2 Exploration Lab, 25-29 June, 2018.

The STARRY project has received funding from
the European Union's Horizon 2020 research and innovation programme
under MSCA ITN-EID grant agreement No 676036.

I. Mendigut\'ia acknowledges the funds from a `Talento' Fellowship (2016-T1/TIC-1890, Government of Comunidad Autónoma de Madrid, Spain).

This work has made use of data from the European Space Agency (ESA) mission
{\it Gaia} (\url{https://www.cosmos.esa.int/gaia}), processed by the {\it Gaia}
Data Processing and Analysis Consortium (DPAC,
\url{https://www.cosmos.esa.int/web/gaia/dpac/consortium}). Funding for the DPAC
has been provided by national institutions, in particular the institutions
participating in the {\it Gaia} Multilateral Agreement.

This publication has made use of the BeSS database, operated at LESIA, Observatoire de Meudon, France: http://basebe.obspm.fr.

This research has made use of data products from the Wide-field Infrared Survey Explorer, which is a joint project of the University of California, Los Angeles, and the Jet Propulsion Laboratory/California Institute of Technology, and NEOWISE, which is a project of the Jet Propulsion Laboratory/California Institute of Technology. WISE and NEOWISE are funded by the National Aeronautics and Space Administration.

This publication has made use of data products from the Two Micron All Sky Survey, which is a joint project of the University of Massachusetts and the Infrared Processing and Analysis Center/California Institute of Technology, funded by the National Aeronautics and Space Administration and the National Science Foundation.

This research has made use of Astropy, a community-developed core Python package for Astronomy \citep{2013A&A...558A..33A, 2018AJ....156..123A}, and the TOPCAT tool (\citealp{2005ASPC..347...29T}). 

In addition, this work has made use of the cross-match service, the VizieR catalogue access tool, the `Aladin sky atlas', and the SIMBAD database developed and operated at CDS, Strasbourg, France.

\end{acknowledgements}


\bibliographystyle{aa} 
\bibliography{AAmybib} 

\begin{appendix}
\section{Disentangling Herbig Ae/Be, CBe stars, and B[e] stars}\label{AppendixA}

In the following we specify which classification decision was made regarding the sources that appear both as Herbig Ae/Be and classical Be in Sect. \ref{Trainingset}. Not all of them have all the observables.

\begin{itemize}
    
\item \object{BD+41 3731} - Classical Be - It appears as a Herbig Ae/Be star in \citet{2013MNRAS.429.1001A} and \citet{2018ApJ...852....5R}. However, \citet{2017AJ....153..252L} consider it a classical Be star and \citet{2014ApJ...797..112C} suggest not to treat it as a PMS object and so did we.

    \item \object{GU CMA} - Herbig Ae/Be - It is generally considered as a Herbig Ae/Be star (e.g. \citealp{2017MNRAS.472..854A}; \citealp{2015MNRAS.453..976F}; \citealp{2018ApJ...852....5R}; \citealp{2018ApJ...857...30M}).

    \item \object{HBC 7} - Herbig Ae/Be - It is a bit doubtful but \citet{2004AJ....127.1682H} argue that it shows charecteristics of PMS objects.

    \item \object{HD 114981} - Classical Be - It appears as Herbig Ae/Be in many papers (\citealp{2018ApJ...852....5R}; \citealp{2015MNRAS.453..976F}) but as CBe in \citet{2017AJ....153..252L}. \citet{2014ApJ...797..112C} found evidence for it to be a CBe star.

    \item \object{HD 130437} - Classical Be - Although it appears in \citet{1994A&AS..104..315T} as a Herbig Ae/Be star, the situation is very unclear. We decided to follow the intuition of \citet{2006A&A...457..171A}.

    \item \object{HD 158643} - Herbig Ae/Be - \citet{2018A&A...609A.108S}

    \item \object{HD 174571} - Herbig Ae/Be - It displays a doubtful nature in many papers (\citealp{2018ApJ...852....5R}; \citealp{2017MNRAS.472..854A}; \citealp{2014ApJ...797..112C}; \citealp{2017AJ....153..252L}) but there is a general consensus that it is a Herbig Ae/Be star.

    \item \object{HD 36408} - Herbig Ae/Be - \citet{2011AJ....141...46D} and \citet{2014ApJ...797..112C}

    \item \object{HD 37490} - Classical Be - \citet{2014ApJ...797..112C} and \citet{2019A&A...621A.123C}

    \item \object{HD 50083} - Herbig Ae/Be - \citet{2018ApJ...852....5R},  \citet{2013MNRAS.429.1001A}, \citet{2010AJ....139...27S}, \citet{2010MNRAS.401.1199W}, and \citet{2014ApJ...797..112C}

    \item \object{HD 76534} - Herbig Ae/Be - \citet{2017ApJ...836..214P}

    \item \object{HD 94509} - Herbig Ae/Be - \citet{2015MNRAS.453..976F}

    \item \object{LkHA 350} - Herbig Ae/Be -  \citet{2004AJ....127.1682H}

    \item \object{MWC 655} - Herbig Ae/Be - \citet{2017MNRAS.472..854A} and \citet{2010MNRAS.401.1199W}
    
    \item \object{V1493 Cyg} - Herbig Ae/Be - It has been little studied in the recent years but appears as a Herbig Ae/Be star in \citet{1994A&AS..104..315T} and in a few papers since then (e.g. \citealp{2018ApJ...857...30M}) although \citet{2004AJ....127.1682H} was unable to classify it.
        
\end{itemize}

Regarding the unclassified B[e] or FS CMa stars, almost all the confirmed FS CMa objects are listed in \citet{1998A&A...340..117L}, \citet{2007ApJ...667..497M}, \citet{2007ApJ...671..828M}, \citet{2017ASPC..508..387M}, and \citet{2018ApJ...856..158K} and they add up to 53 objects (around 70 proposed in total, \citealp{2015EAS....71..181M}). A total of 17 FS CMa stars from this list were discarded from the sets of known PMS and CBe stars:

\begin{itemize}

\item \object{BD+23 3183}
\item \object{CD-24 5721}
\item \object{CD-49 3441}
\item \object{AS  119}
\item \object{HD 328990}
\item \object{HD 45677}
\item \object{HD 50138}
\item \object{HD 85567}
\item \object{Hen 3-847}
\item \object{LkHA  348}
\item \object{MWC 1055}
\item \object{MWC 342}
\item \object{MWC 657}
\item \object{PDS 021}
\item \object{PDS 211}
\item \object{V2211 Cyg}
\item \object{V669 Cep}
\end{itemize}

Separately, but related, in \citet{2018A&A...620A.128V} it was found that because of their positions on the HR diagram: \object{MWC 314}, \object{MWC 623}, and \object{MWC 930} were not very likely to be PMS objects. Indeed, \object{MWC 314} seems to be a supergiant B[e] star (\citealp{2016A&A...585A..60F}), \object{MWC 930} looks like a luminous blue variable (\citealp{2016A&A...587A.115M}; \citealp{2018Natur.561..498J}) and \object{MWC 623} seems clear to be a FS CMa star (\citealp{2007ApJ...667..497M}; \citealp{2018A&A...617A..79P}). Therefore, we also removed these three objects from our set of known HAeBes (Sect.\ref{PMS}).

\section{Algorithm and methodology}\label{S_Description of the pipeline}

\begin{figure}[ht!]
\includegraphics[scale=0.564]{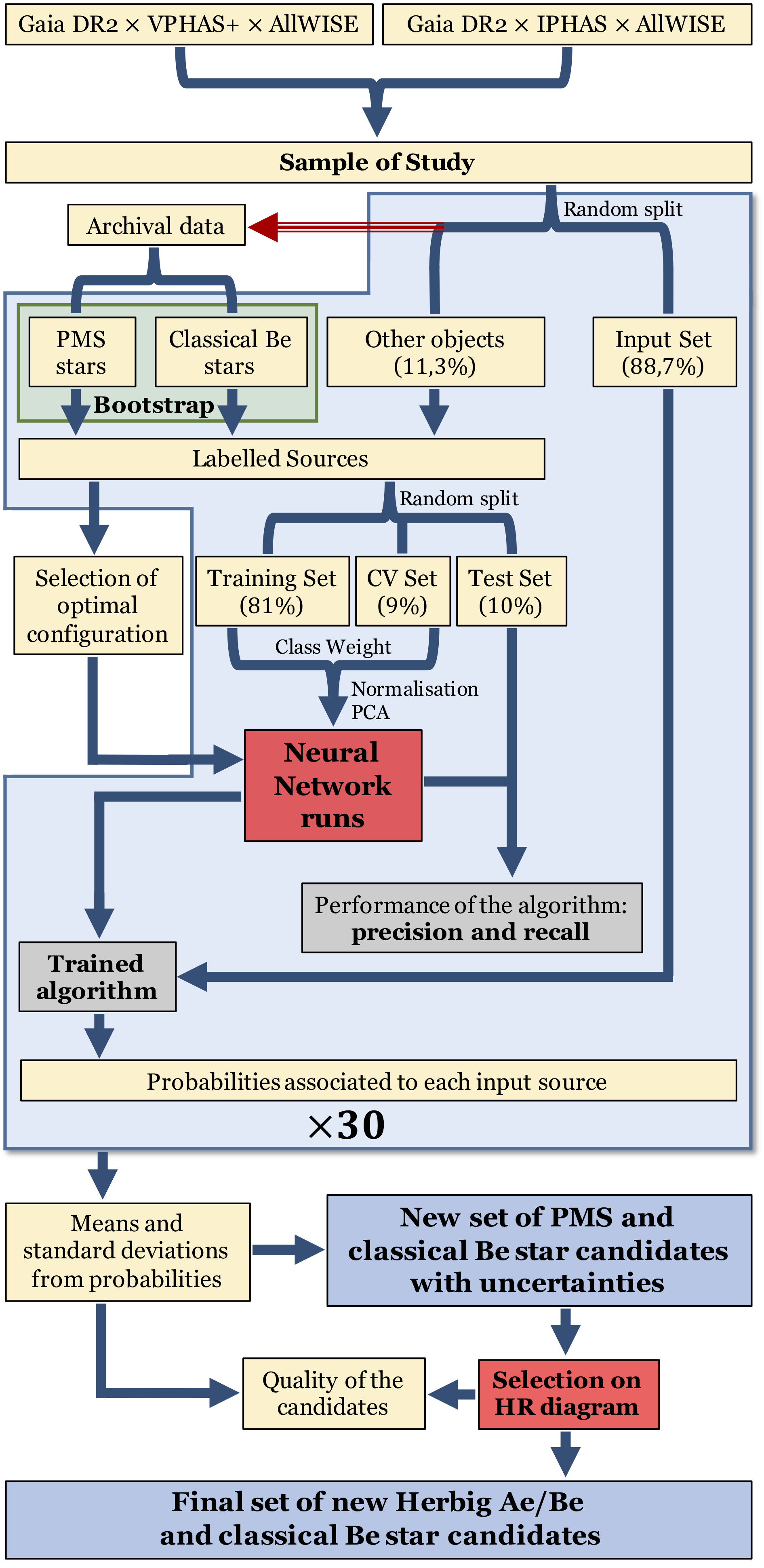}
\caption{Pipeline of the whole algorithm, from the cross-match of \textit{Gaia} DR2, AllWISE, IPHAS, and VPHAS+ to the set of new HAeBes and classical Be stars. The light blue area indicates the set of processes that are repeated in a loop 30 times, each time generating a different set of probabilities associated to each input source. The green area shows the bootstrapped sets. The red arrow indicates that the Archival data is partially contained within the Sample of Study, but was also constructed using external information like the H$\alpha$ EWs.}\label{fig:Pipeline2}
\end{figure}

The pipeline used can be seen in Fig. \ref{fig:Pipeline2}. Most of the algorithm described in it is available on a GitHub repository \footnote{\href{https://github.com/MVioque/YODA}{https://github.com/MVioque/YODA}} under the name of YODA (Young Object Discoverer Algorithm). In this appendix we describe this pipeline in detail.

\subsection{Class weights}\label{Weights}

As explained in Sect. \ref{Other_sources}, the set of other objects is contaminated by undiscovered PMS and CBe stars. This causes the ANN used in Sect \ref{Other_sources} to obtain a decent precision but a terrible recall (see Fig. \ref{fig:size_other}). As discussed in Sect. \ref{Other_sources}, this is because we are only retrieving the less common sources. Any pre-classifications performed with the observables used by the algorithm would artificially bias the results. One option is to use simulations to generate this well defined category of other objects without PMS and CBe stars (as done for example by \citealp{2018A&A...618A..59C}) but there is none that lists PMS and CBe objects and contains IR and H$\alpha$ information. 

We can address this issue by changing the weights of the sources used in the training, in a way that they are balanced for the different category sizes. Hence, during training the ANN is heavily penalized when failing at categorising PMS or CBe sources, but lightly affected by mistakes on the other objects category, which is much larger. Therefore, the training is not dominated by contaminants or undiscovered sources, although they still are considered. This weighting technique produces a decent recall and a very low precision, but this precision is just a lower limit as the candidate regions of the feature space contain many undiscovered PMS and CBe stars. In other words, FP is over-measured (Eq. \ref{E_Precision}). In addition, this class weights stress the algorithm to focus on the differences between PMS sources and CBe stars, which also bias the selection of PMS sources towards the high-mass end. 

\subsection{Architecture selection}\label{architecture}

Different Machine Learning algorithms were considered for this classification problem. A variety of them have been used so far for similar matters. For example: random forests (\citealp{2018MNRAS.476.2968H}; \citealp{2019MNRAS.487.2522M}; \citealp{2019A&A...625A..97R}), support vector machines (\citealp{2013A&A...557A..16M}; \citealp{2016MNRAS.458.3479M}; \citealp{2017A&A...606A..39S}; \citealp{2018MNRAS.479.2389K}), or artificial neural networks (\citealp{2001ApJ...562..528S}; \citealp{2017MNRAS.470.3395H}). However, similar performances are achieved with most of the algorithms and it is evident that the output is mainly dominated by the quality of the training data (\citealp{2017A&A...605A.123P}; \citealp{2018MNRAS.475.2326P}, or \citealp{2019MNRAS.487.2522M} in a similar problem of identifying Young-Stellar Objects). Therefore, we decided to use a shallow artificial neural network as it has the advantage of flexibility and non-linearity, being able to describe very complex and subtle relations. In addition, its output can be a probability vector, which eases the catalogue construction. Cons are the number of hyper-parameters required, which are normally hard to interpret. 

Therefore, we needed to find the architecture or optimal configuration of the ANN for our particular problem. This means choosing the hyper-parameters of the ANN (e.g. layers, neurons per layer, regularization). Ideally, this architecture would be selected with cross-validation (CV) sets that are not used for the training. However, our sample of known PMS and CBe stars is too small to have the number of CV sets necessary to test a large enough grid of ANN configurations. Instead, we used the set of Labelled Sources to select the optimal hyper-parameters and then, independently, used those hyper-parameters and that same set of Labelled Sources for the training (see Fig. \ref{fig:Pipeline2}). We ran an ANN over the Labelled Sources set 100 times (each time with a 10\% test set random split); evaluating at each training iteration on a CV set (sized 10\%) and early-stopping whenever the precision of the algorithm on the PMS category (selecting as candidates those with a probability $p\geq50\%$ of belonging to such category) stopped increasing over 250 iterations. In addition, we imposed that the recall had to be at least 90\%. In each run a different grid of hyper-parameters was used. After the 100 runs, the best architecture resulted in two fully connected hidden layers of 580 neurons, with a dropout rate of 50\% and $0.01$ L2 regularization. Batch Normalisation was applied after every layer, though no batches were used and the whole training data was evaluated in each training iteration (as it is a very skewed training set, see Sect. \ref{Trainingset}).

The activation functions used were `ReLU' for the hidden layers and `softmax' for the output layer. This is because softmax output can be interpreted as a probability distribution. The loss function used was `categorical crossentropy' with the `AdaMax' optimizer (\citealp{2014arXiv1412.6980K}). To construct the ANNs of this project we used Keras (\citealp{chollet2015keras}), a high-level neural networks application programming interface.

\subsection{Training, cross-validation, and test set}\label{Traing_Test}

We shuffle the Labelled Sources set and randomly split it into two subsets (see Fig. \ref{fig:Pipeline2}). One contains $90\%$ of the sources and is used to train the algorithm (training set). The other, containing $10\%$ of the sources is used to evaluate its performance (test set).

The first step is to perform feature scaling and mean normalisation to the observables, so they all have the same mean and standard deviation. Then we apply PCA to the scaled observables to get the set of features used by the ANN (12 principal components of the 48 carry $99.99\%$ of the variance, see Sect. \ref{S_parameters}). Next, we train the ANN, which has the architecture chosen in Appx. \ref{architecture}, with the training set and use a CV set (sized $10\%$ of the training set) to evaluate the ANN performance after every training iteration. Early-stopping finishes the training whenever the precision on the PMS category (with $p\geq50\%$) stops increasing over 50 iterations. We note that, as discussed in Sect. \ref{Weights}, the precision retrieved is just a lower limit. Once the ANN is trained, we run it on the test set, that needs to be scaled and feature extracted as done for the training set. Evaluation on test set gives a value of precision and recall for each probability threshold for classification $p$ (i.e. the performance of the algorithm, see Fig. \ref{PrevsRe}). Finally, we can apply the trained ANN to the Input Set, giving a probability for every source of belonging to each of the chosen categories.

\subsection{Bootstrap}\label{Bootstrap}

\begin{figure}[t]
\includegraphics[scale=0.448]{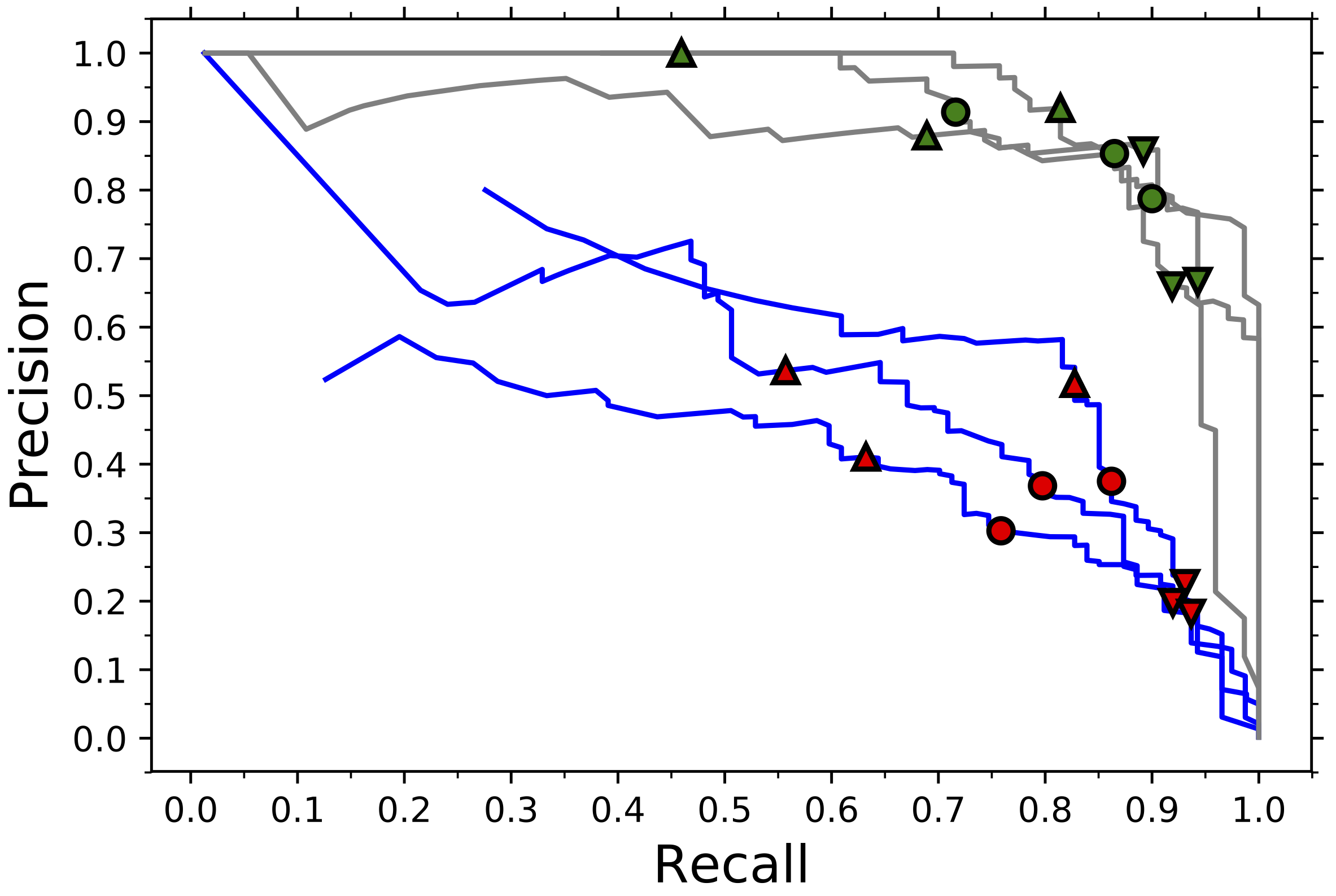}
\caption{Precision vs. recall trade-off plot resulting after evaluation on test set for three different bootstrapped iterations. Blue lines correspond to PMS classification and grey lines to classical Be classification. Different probability thresholds ($p$) for selecting candidate objects correspond to different locations on the line. Upwards arrows ($p\geq75\%$), circles ($p\geq50\%$), and downwards arrows ($p\geq25\%$) are examples of such probability thresholds. Some lines do not cover the whole metric space because evaluation stops whenever there are no longer true positives in the corresponding test set. The precision values are lower limits to the real precisions.\label{PrevsRe}}
\end{figure} 

A major issue is the small size of the PMS and classical Be categories in the training data (see Sect. \ref{Trainingset}). Small training sets imply that outliers and contaminants have a very strong influence and might dominate the posterior generalisation. In addition, the training might be biased to any hidden trend or pattern.

One way to minimise the impact of this is by means of the bootstrap. The key idea is to fake the construction of new training sets. It works by repeatedly sampling the original training data and randomly substituting sources with others of the same data set. If we run the same algorithm over two bootstrapped sets we obtain similar, but slightly different metrics as a result. If we repeat this bootstrapping process a large enough number of times we end up with a distribution of precisions and recalls characteristic of our method, which allows us to estimate the uncertainty of the metrics for each probability threshold. Bootstrapping has another advantage, which is to better represent the distribution of the categories on the feature space and minimise the impact of outliers.

Therefore, we run the processes described in Appx. \ref{Traing_Test} (blue area of Fig. \ref{fig:Pipeline2}) 30 times in a loop. In each iteration, we create a bootstrapped version of the combination of the categories of known PMS and classical Be stars (so the number of objects in each group is not conserved). In the case of the other objects category, we just withdraw another random set of sources from the Sample of Study. Once the algorithm is trained with a certain Labelled Sources bootstrapped set, we obtain by evaluating on the corresponding test set values for the precision and recall at different probability thresholds (see Fig. \ref{PrevsRe}). When we run the trained ANN over the Input Set, we retrieve probabilities associated to every source of belonging to each of the three categories. Hence, after the bootstrapped iterations we end up with 30 values for precision and recall at different thresholds (in Fig. \ref{PrevsRe} only three bootstrapped iterations are shown for clarity) and 30 sets of probabilities associated to each source of the Sample of Study. This is because as the category of other objects has been randomly sampled 30 times, the whole SoSt has been covered eventually. To obtain the final values presented in Tables \ref{Table2} and \ref{Table3} we average these 30 repetitions and take the standard deviation of the mean as the uncertainty of each measurement.

\section{Visualisation}\label{S_Visualization}
\begin{figure*}[h!]
\includegraphics[scale=0.63]{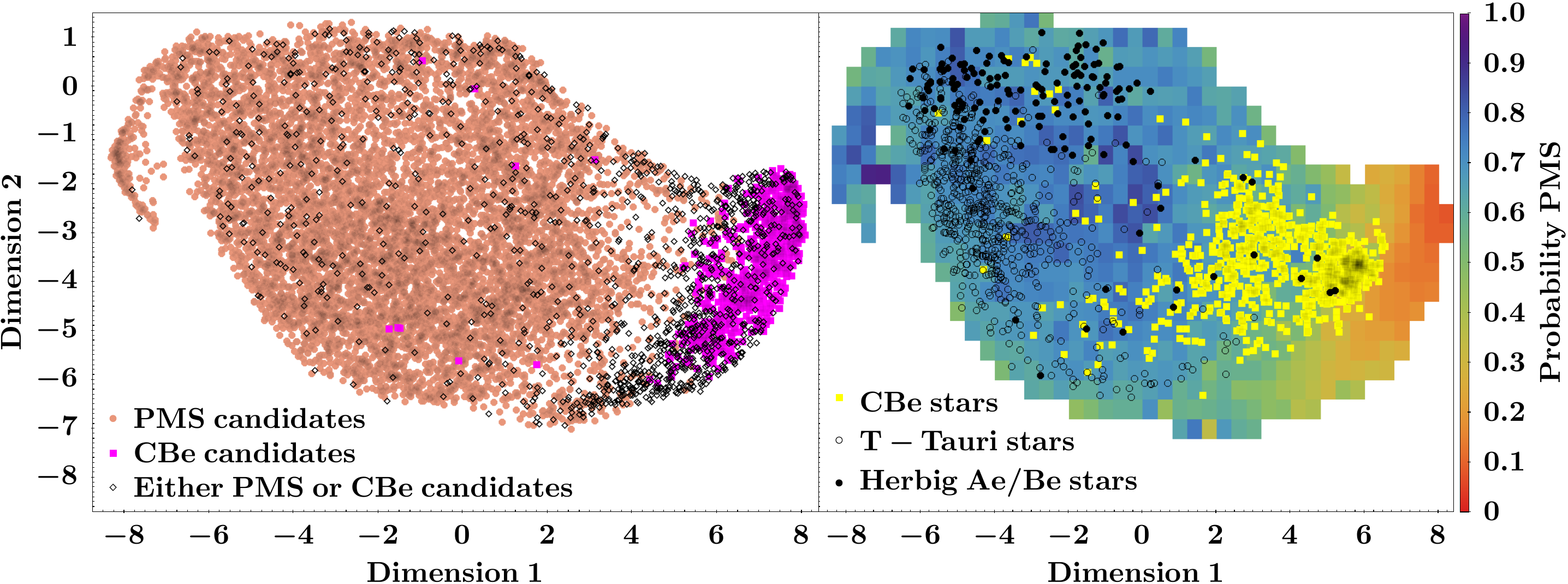}
\caption{UMAP dimensionality reduction from the 12D space of features to 2D. \textit{Left:} dimensionality reduction of the PMS and classical Be candidates, together with those sources which belong to either category (i.e. $p(PMS)+p(CBe)\geq50\%$ but $p(PMS)<50\%$, $p(CBe)<50\%$). \textit{Right:} we project the known Herbig Ae/Be, T-Tauri, and classical Be stars used for the training onto the same plane, which is colour-coded following the PMS probability distribution of the sources at left.}\label{projection}
\end{figure*}

It is possible to visualise the feature space and the selection using a dimensionality reduction algorithm. We used the UMAP algorithm (Uniform Manifold
Approximation and Projection for
Dimension Reduction, \citealp{2018arXiv180203426M}) to project the 12-dimensional feature space (Sect. \ref{PCA}) into two dimensions. This is done in Fig. \ref{projection}. At left we project the candidates using an euclidean metric (15\% of the number of sources as number of neighbours and a minimum distance of $0.4$). In Fig. \ref{projection} at right we project the known HAeBe, T-Tauri, and CBe stars used for the training (Sect. \ref{Trainingset}) onto this same plane, which is colour-coded following the PMS probability distribution of the sources at left. This dimensionality reduction helps to understand the catalogue construction and to find trends within the data. However, information is lost when moving to 2D. The category of other objects was not included here because of size limitations.

First, we can see that there is indeed a separation between known PMS sources and classical Be stars (Fig. \ref{projection} at right), which is used by the algorithm to learn how to separate these populations (Fig. \ref{projection} at left). It is remarkable that most of the retrieved CBe candidates are not found where known CBe stars are, but even farther away from the PMS region, which might imply that we are retrieving very extreme CBe candidates (see Sect. \ref{S_Evaluation}). In addition, most of the PMS candidates that are located very close to the CBe region are those with a high CBe probability and vice-versa (see Fig. \ref{fig:Probability_map}). These PMS candidates have $r-H\alpha$ and $G_{BP}-G_{RP}$ values typical of known CBes and low IR-excesses for PMS objects. However, these are still typically larger than the ones of known CBes. This, together with strong $G_{var}$ variability explains their selection as PMS. 

Second, in Fig. \ref{projection} an arm structure at the very top left of the space of PMS candidates seems special. A closer look at the 204 PMS candidates located in that arm shows that they are all placed in the red giant region of the HR diagram (see Fig. \ref{HR_grid} at top left). They differentiate from the rest of the PMS candidates in that they have more variability in our indicators and are typically brighter. In addition, they show on average larger near-IR excesses but lower mid-IR excesses. None of them have reliable astrometry so they do not contaminate our sample of Herbig Ae/Be candidates. Potentially, they are evolved contaminants and are flagged in Table \ref{Table2} as `G\textsubscript{UMAP}' (see Sect. \ref{S_Evaluation} for further details of their nature). We cannot exclude sources only because of their HR diagram position, as we might remove HAeBes with high extinctions and some candidates do not have parallax information. Hence, candidates in similar red giant HR diagram positions but not in this UMAP region are not flagged in the final catalogues of Tables \ref{Table2} and \ref{Table3}.
\newpage

\section{Catalogue of new PMS and classical Be stars}\label{AppendixD}

Here we present a portion of the catalogue of new pre-main sequence (Table \ref{Table2}) and classical Be (Table \ref{Table3}) stars for guidance regarding its form and content. These tables are available in electronic form in their entirety with uncertainties of the magnitudes, quality flags and rest of \textit{Gaia} parameters together with angular distances from AllWISE and IPHAS or VPHAS+ sources to \textit{Gaia} sources. Below a description of the possible warning flags of Tables \ref{Table2} and \ref{Table3} by alphabetical order. See the main text for further discussion.

\begin{itemize}
    \item G\textsubscript{UMAP}: Possible evolved star contaminant. Identified through UMAP visualisation. Discussed in Appx. \ref{S_Visualization}.
    \item ID AllW: Source with an AllWISE name repeated in the Sample of Study. Discussed in Sect. \ref{S_Evaluation}, point 3.
    \item ID IPH/VPH: Source with an IPHAS or VPHAS+ name repeated in the Sample of Study. Discussed in Sect. \ref{S_Evaluation}, point 3.
    \item PN: Possible Planetary Nebula or `unclassified B[e]' contaminant. Defined as those candidates with $r-H\alpha\geq1.3$. Discussed in Sect. \ref{S_Evaluation}, point 5.
    \item Var: Photometrically variable PMS candidate. Defined as those PMS candidates with $G_{var}\geq10$. Discussed in Sect. \ref{S_Variability}.
    \item W3W4: Source which extended source flag of AllWISE catalogue is different of 0. Discussed in Sect. \ref{S_Evaluation}, point 2.
    \item X-mtch: Likely false candidate because of incorrect cross-match with IPHAS or VPHAS+. Discussed in Sect. \ref{S_out}.
\end{itemize}

\newpage
\begin{table*}
\caption{PMS candidates ($p\geq50\%$, 8470 sources) ordered by probability. In boldface sources that are strong Herbig Ae/Be candidates according to their position on the HR diagram ($M_{G}<6$, 1361, see Fig. \ref{Progress}). These Herbig Ae/Be candidates are also indicated in the column `HAeBe'.}              
\label{Table2}      
\setlength{\tabcolsep}{5pt}
\centering                                      
\begin{tabular}{l c c c c c c c}          
\hline\hline
\addlinespace[0.07cm]
\textit{Gaia} source id & RA & DEC & Probability & Probability & Probability & V\textsubscript{htg} & G\textsubscript{var} \\ 
& h:m:s & deg:m:s & PMS & CBe & Other &  &\\ 
\hline
   \addlinespace[0.07cm]
   \textbf{\object{189190423171051136}} & \textbf{05:39:18.1} & \textbf{+36:17:16} & \textbf{1.00$\pm$0.00} & \bm{$(7.8\pm2.0)\cdot10^{-6}$} & \bm{$(4.2\pm1.6)\cdot10^{-6}$} & \textbf{-52.13} & \textbf{158.74}\\
   \object{513068993519575808} & 01:39:32.5 & +64:53:02 & 1.00$\pm$0.00 & $(4.18\pm0.91)\cdot10^{-5}$ & $(7.7\pm2.9)\cdot10^{-6}$ & -66.16 & 165.17\\   
   \object{5546522453993928960} & 08:12:40.5 & -34:14:12 & 1.00$\pm$0.00 & $(4.4\pm2.0)\cdot10^{-5}$ & $(1.38\pm0.58)\cdot10^{-5}$ & -117.08 & 225.56\\    
   \object{181175743617347328} & 05:23:04.3 & +33:28:46 & 1.00$\pm$0.00 & $(1.67\pm0.44)\cdot10^{-4}$ & $(4.7\pm2.0)\cdot10^{-6}$ & -47.67 & 131.31\\ 
   
   \textbf{\object{3430718965791698176}} & \textbf{05:52:51.7} & \textbf{+26:47:30} & \textbf{1.00$\pm$0.00} & \bm{$(3.04\pm0.96)\cdot10^{-4}$} & \bm{$(3.6\pm1.0)\cdot10^{-5}$} & \textbf{-67.90} & \textbf{152.91}\\ 

   \object{3046391307734381184} & 07:09:22.3 & -10:30:57 & 1.00$\pm$0.00 & $(2.06\pm0.55)\cdot10^{-4}$ & $(1.52\pm0.57)\cdot10^{-4}$ & -44.33 & 77.08\\
   
   \object{3449189833426211840} & 05:32:13.9 & +34:06:01 & 1.00$\pm$0.00 & $(8.0\pm2.8)\cdot10^{-5}$ & $(2.8\pm1.8)\cdot10^{-4}$ & -107.39 & 183.40\\ 

   \object{189577661722195840} & 05:37:53.2 & +37:24:56 & 1.00$\pm$0.00 & $(2.61\pm0.60)\cdot10^{-4}$ & $(1.52\pm0.51)\cdot10^{-4}$ & -66.15 & 117.40\\ 

   \object{2006046771487587328} & 22:24:13.8 & +56:11:33 & 1.00$\pm$0.00 & $(2.3\pm1.1)\cdot10^{-4}$ & $(2.38\pm0.99)\cdot10^{-4}$ & -117.72 & 162.47\\ 
   \textbf{\object{4279295067717293696}} & \textbf{18:47:32.6} & \textbf{+02:12:06} & \textbf{1.00$\pm$0.00} & \bm{$(1.31\pm0.21)\cdot10^{-4}$} & \bm{$(3.82\pm0.94)\cdot10^{-4}$} & \textbf{-33.04} & \textbf{96.61}\\

   \object{3114712967419900544} & 07:09:42.8 & +01:53:11 & 1.00$\pm$0.00 & $(5.5\pm2.4)\cdot10^{-5}$ & $(5.4\pm4.0)\cdot10^{-4}$ & -46.52 & 143.75\\ 
   
   \object{3102576519414606464} & 06:58:34.4 & -03:56:46 & 1.00$\pm$0.00 & $(1.75\pm0.58)\cdot10^{-4}$ & $(4.3\pm1.2)\cdot10^{-4}$ & -26.55 & 58.08\\
   
   ... & ... & ... & ... & ... & ... & ... & ... \\    
\hline      
\end{tabular}
\end{table*}
\begin{table*}
\setlength{\tabcolsep}{5pt}
\centering                                      
\begin{tabular}{c c c c c c c c c c c c c}          
\hline\hline
\addlinespace[0.07cm]
G\textsubscript{BP} & G & G\textsubscript{RP} & IPHAS or VPHAS+ & r & i & H$\alpha$  & AllWISE & J & H & K\textsubscript{s}\\
(mag) & (mag) & (mag) & name & (mag) & (mag) & (mag) & name & (mag) & (mag) & (mag)\\ 
\hline
   \addlinespace[0.07cm]
   \textbf{15.97} & \textbf{15.04} & \textbf{14.04} & \textbf{J053918.09+361716.2} & \textbf{13.78} & \textbf{13.06} & \textbf{13.21} & \textbf{J053918.08+361716.1} & \textbf{12.98} & \textbf{11.23} & \textbf{9.79}\\  
   14.34 & 13.80 & 12.98 & J013932.55+645302.3 & 12.91 & 12.43 & 12.35 & J013932.53+645302.2 & 11.43 & 10.21 & 9.11\\
   14.97 & 14.18 & 13.20 & J081240.5-341411.7 & 12.86 & 12.19 & 11.99 & J081240.50-341411.7 & 10.86 & 10.08 & 9.59\\
   14.89 & 14.40 & 13.66 & J052304.26+332846.5 & 13.80 & 13.24 & 12.90 & J052304.26+332846.4 & 12.01 & 11.12 & 10.25\\ 
   
   \textbf{14.40} & \textbf{13.94} & \textbf{13.05} & \textbf{J055251.75+264730.2} & \textbf{12.62} & \textbf{12.22} & \textbf{12.28} & \textbf{J055251.74+264730.0} & \textbf{12.16} & \textbf{11.04} & \textbf{10.06}\\    
   
   16.75 & 15.56 & 14.31 & J070922.3-103057.0 & 16.04 & 14.94 & 14.78 & J070922.28-103057.0 & 12.47 & 10.98 & 9.90\\ 
   
   16.17 & 14.76 & 13.63 & J053213.92+340601.5 & 13.70 & 12.39 & 13.25 & J053213.92+340601.4 & 10.80 & 9.83 & 8.83\\    
   
   14.73 & 13.68 & 12.63 & J053753.15+372456.1 & 14.30 & 13.07 & 13.80 & J053753.15+372456.0 & 10.54 & 9.83 & 9.31\\    
   
   14.00 & 13.31 & 12.40 & J222413.78+561133.2 & 13.63 & 12.92 & 13.29 & J222413.75+561133.2 & 9.56 & 8.69 & 8.01\\    
   
   \textbf{15.15} & \textbf{14.11} & \textbf{13.05} & \textbf{J184732.62+021205.9} & \textbf{15.01} & \textbf{13.84} & \textbf{14.23} & \textbf{J184732.61+021205.7} & \textbf{11.34} & \textbf{9.97} & \textbf{8.79}\\
   
   13.62 & 11.79 & 10.19 & J070942.85+015311.0 & 13.86 & 11.84 & 13.15 & J070942.75+015311.1 & 8.20 & 6.62 & 5.25\\     

   18.24 & 17.57 & 16.41 &  J065834.4-035645.6 & 21.41 & 20.84 & 20.13 & J065834.39-035645.6 & 13.73 & 12.40 & 11.38\\  
   
   ... & ... & ... & ...  & ... & ... & ... & ... & ... & ... & ... \\     
\hline      
\end{tabular}
\end{table*}
\begin{table*}
\setlength{\tabcolsep}{4.3pt}
\centering                                      
\begin{tabular}{c c c c c c c c c c c c}          
\hline\hline
\addlinespace[0.07cm]
W1 & W2 & W3 & W4 & RUWE & Parallax ($\varpi$) & Distance & $A_{G}'$ & M\textsubscript{G} & G\textsubscript{BP}-G\textsubscript{RP} & HAeBe & Flag\\
(mag) & (mag) & (mag) & (mag) &  & [mas] & [pc] & (mag) & (mag) & (mag) & & \\ 
\hline 
   \addlinespace[0.07cm]
   \textbf{7.90} & \textbf{7.05} & \textbf{4.84} & \textbf{3.23} & \textbf{1.36} & \bm{$0.636\pm0.048$} & \bm{$1510^{+120}_{-100}$} & \textbf{1.06} & \textbf{3.09} & \textbf{1.41} & \textbf{Yes} & \textbf{Var}\\
   \addlinespace[0.07cm]
   7.95 & 7.26 & 5.41 & 3.90 & 1.86 & $1.116\pm0.029$ & $874^{+23}_{-22}$ & - & - & - & - & Var\\ 
   \addlinespace[0.07cm]
   8.63 & 7.81 & 5.37 & 3.32 & 1.33 & $2.877\pm0.025$ & $344.2^{+3.0}_{-2.9}$ & 0.06 & 6.43 & 1.75 & - & Var, W3W4\\
   \addlinespace[0.07cm]
   9.12 & 8.39 & 6.35 & 4.21 & 1.37 & $0.234\pm0.040$ & $3710^{+650}_{-490}$ & 1.04 & 0.51 & 0.72 & - & Var, W3W4\\ 
   
   \addlinespace[0.07cm]
   \textbf{9.26} & \textbf{8.76} & \textbf{7.23} & \textbf{5.56} & \textbf{1.09} & \bm{$1.950\pm0.032$} & \bm{$505.6^{+8.3}_{-8.0}$} & \textbf{0.27} & \textbf{5.15} & \textbf{1.22} & \textbf{Yes} & \textbf{Var}\\   
   
   \addlinespace[0.07cm]
   7.19 & 5.90 & 3.19 & 1.06 & 3.28 & $1.00\pm0.20$ & $1020^{+320}_{-200}$ & - & - & - & - & Var, W3W4\\   
   
   \addlinespace[0.07cm]
   6.92 & 5.79 & 3.95 & 2.94 & 1.01 & $-0.068\pm0.080$ & $7200^{+2500}_{-1800}$ & - & - & - & - & G\textsubscript{UMAP}, Var\\   
   
   \addlinespace[0.07cm]
   8.34 & 7.37 & 5.01 & 3.66 & 1.32 & $0.025\pm0.041$ & $8200^{+2300}_{-1600}$ & - & - & - & - & Var, W3W4\\   
   
   \addlinespace[0.07cm]
   7.06 & 5.99 & 3.79 & 2.43 & 1.72 & $0.326\pm0.027$ & $2810^{+230}_{-200}$ & - & - & - & - & Var, W3W4\\   
   
   \addlinespace[0.07cm]
   \textbf{7.61} & \textbf{6.97} & \textbf{4.84} & \textbf{3.10} & \textbf{1.30} & \bm{$1.614\pm0.046$} & \bm{$609^{+18}_{-17}$} & \textbf{1.99} & \textbf{3.19} & \textbf{1.13} & \textbf{Yes} & \textbf{Var}\\
   
   \addlinespace[0.07cm]
   3.99 & 1.79 & 1.93 & 1.44 & 1.01 & $0.103\pm0.089$ & $4500^{+1600}_{-1100}$ & - & - & - & - & G\textsubscript{UMAP}, Var, W3W4\\   
   
   \addlinespace[0.07cm]
   10.58 & 9.25 & 6.40 & 4.02 & 1.15 & $0.36\pm0.11$ & $2580^{+1120}_{-640}$ & - & - & - & - & Var, W3W4\\   
  
   ... & ... & ... & ... & ... & ... & ... & ... & ... & ... & ... & ...\\    
\hline      
\end{tabular}
\tablefoot{This table is available in electronic form in its entirety with uncertainties of the magnitudes, quality flags, and rest of \textit{Gaia} parameters together with angular distances from AllWISE and IPHAS or VPHAS+ sources to \textit{Gaia} sources. A portion is shown here for guidance regarding its form and content. The probabilities are expressed to the precision of their uncertainties. Distances from \citet{2018AJ....156...58B}. $A_{G}'$ only traces the interstellar extinction. It is used to correct $M_{G}$ and $G_{BP}-G_{RP}$. $A_{G}'$, $M_{G}$, and $G_{BP}-G_{RP}$ are only presented for sources with RUWE<1.4 and $\varpi/\sigma(\varpi)\geq5$. By construction Herbig Ae/Be candidates are astrometrically well behaved (RUWE<1.4 and $\varpi/\sigma(\varpi)\geq10$). These constraints can be relaxed to obtain more Herbig Ae/Be candidates. The probabilities of the three categories sum up to 1. The different catalogue warning flags are discussed throughout the text and summarised in Appendix \ref{AppendixD}.}
\end{table*}

\clearpage

\begin{table*}
\caption{A representative sample of the full table of classical Be candidates ($p\geq50\%$, 693 sources) ordered by probability.}   
\label{Table3}      
\setlength{\tabcolsep}{5pt}
\centering                                      
\begin{tabular}{l c c c c c c c}          
\hline\hline
\addlinespace[0.07cm]
\textit{Gaia} source id & RA & DEC & Probability & Probability & Probability & V\textsubscript{htg} & G\textsubscript{var} \\ 
& h:m:s & deg:m:s & PMS & CBe & Other &  &\\ 
\hline 
   ... & ... & ... & ... & ... & ... & ... & ... \\ 
   \object{2012831922144678272} & 23:43:35.7 & +61:27:44 & 0.119$\pm$0.010 & 0.831$\pm$0.015 & 0.0500$\pm$0.0061 & -1.87 & 7.75\\
   
   \object{3106114885277650944} & 06:49:27.0 & -02:33:30 & 0.105$\pm$0.014 & 0.831$\pm$0.020 & 0.064$\pm$0.010 & -1.36 & 2.40\\ 
   
   \object{509370275062324608} & 01:34:15.6 & +59:30:59 & 0.146$\pm$0.012 & 0.830$\pm$0.014 & 0.0239$\pm$0.0032 & -7.87 & 13.95\\  
   
   \object{188951970886326656} & 05:11:53.5 & +40:13:11 & 0.1127$\pm$0.0099 & 0.830$\pm$0.015 & 0.0572$\pm$0.0081 & 0.14 & 1.62\\ 
   
   \object{2173867430250308608} & 21:50:31.8 & +54:04:53 & 0.122$\pm$0.016 & 0.830$\pm$0.019 & 0.0483$\pm$0.0058 & -10.14 & 15.16\\ 
   
   \object{4268987932187374848} & 19:07:10.6 & +03:29:22 & 0.107$\pm$0.010 & 0.828$\pm$0.017 & 0.0644$\pm$0.0090 & -0.42 & 6.07\\
   
   \object{518388499496394752} & 01:58:16.4 & +65:49:52 & 0.107$\pm$0.010 & 0.828$\pm$0.015 & 0.0647$\pm$0.0071 & -7.34 & 11.67\\
   
   \object{2201241387128842240} & 22:18:59.0 & +59:47:10 & 0.137$\pm$0.018 & 0.827$\pm$0.022 & 0.0356$\pm$0.0083 & 0.00 & 0.67\\  
   
   \object{2060197062012903296} & 20:00:28.0 & +37:06:52 & 0.103$\pm$0.012 & 0.827$\pm$0.017 & 0.0700$\pm$0.0080 & 0.08 & 1.18\\ 
   
   \object{3325867711602501504} & 06:21:13.0 & +09:07:21 & 0.101$\pm$0.011 & 0.827$\pm$0.019 & 0.072$\pm$0.016 & -0.23 & 0.99\\
   
   \object{3106856785040836224} & 06:37:36.4 & -01:54:03 & 0.107$\pm$0.010 & 0.824$\pm$0.016 & 0.0694$\pm$0.0079 & -6.52 & 12.02\\ 
   
   \object{3356939124228643968} & 06:35:02.5 & +15:17:52 & 0.125$\pm$0.014 & 0.823$\pm$0.017 & 0.0518$\pm$0.0061 & -0.27 & 0.87\\     
   ... & ... & ... & ... & ... & ... & ... & ... \\    
\hline      
\end{tabular}
\end{table*}
\begin{table*}
\setlength{\tabcolsep}{5pt}
\centering                                      
\begin{tabular}{c c c c c c c c c c c c c}          
\hline\hline
\addlinespace[0.07cm]
G\textsubscript{BP} & G & G\textsubscript{RP} & IPHAS or VPHAS+ & r & i & H$\alpha$  & AllWISE & J & H & K\textsubscript{s}\\
(mag) & (mag) & (mag) & name & (mag) & (mag) & (mag) & name & (mag) & (mag) & (mag)\\ 
\hline 
   ... & ... & ... & ...  & ... & ... & ... & ... & ... & ... & ... \\ 
   13.14 & 12.80 & 12.26 & J234335.71+612744.0 & 12.92 & 12.39 & 12.28 & J234335.70+612743.9 & 11.59 & 11.36 & 11.15\\
   
   12.93 & 12.67 & 12.18 & J064926.96-023330.0 & 12.56 & 12.14 & 11.88 & J064926.98-023330.5 & 11.41 & 11.24 & 10.87\\
   
   12.27 & 11.99 & 11.51 & J013415.63+593058.7 & 12.00 & 11.60 & 11.40 & J013415.64+593058.6 & 10.98 & 10.69 & 10.48\\
   
   12.66 & 12.43 & 11.99 & J051153.45+401310.9 & 12.50 & 11.92 & 11.64 & J051153.44+401310.9 & 11.55 & 11.38 & 11.21\\
   
   12.87 & 12.32 & 11.58 & J215031.85+540452.9 & 12.73 & 11.91 & 11.94 & J215031.83+540452.7 & 10.71 & 10.41 & 10.24\\
   
   12.30 & 12.04 & 11.59 & J190710.55+032922.3 & 12.15 & 11.60 & 11.37 & J190710.55+032922.4 & 10.95 & 10.74 & 10.54\\
   
   12.81 & 12.32 & 11.67 & J015816.40+654951.9 & 12.16 & 11.56 & 11.61 & J015816.37+654951.9 & 10.89 & 10.65 & 10.46\\
   
   12.73 & 12.33 & 11.75 & J221859.05+594710.0 & 12.60 & 11.84 & 12.18 & J221859.03+594709.9 & 11.02 & 10.90 & 10.77\\
   
   12.43 & 12.25 & 11.94 & J200028.00+370651.9 & 12.33 & 11.93 & 11.99 & J200028.00+370653.0 & 11.54 & 11.45 & 11.37\\
   
   11.99 & 11.90 & 11.66 & J062112.96+090721.4 & 12.48 & 11.71 & 11.44 & J062112.95+090721.4 & 11.37 & 11.28 & 11.13\\
   
   13.13 & 12.68 & 12.06 & J063736.41-015403.0 & 12.58 & 12.04 & 12.21 & J063736.41-015402.9 & 11.26 & 11.02 & 10.84\\
   
   12.56 & 12.37 & 12.01 & J063502.46+151752.5 & 12.50 & 12.05 & 12.11 & J063502.46+151752.5 & 11.57 & 11.52 & 11.37\\     
   ... & ... & ... & ...  & ... & ... & ... & ... & ... & ... & ... \\     
\hline      
\end{tabular}
\end{table*}
\begin{table*}
\setlength{\tabcolsep}{5pt}
\centering                                      
\begin{tabular}{c c c c c c c c c c c}          
\hline\hline
\addlinespace[0.07cm]
W1 & W2 & W3 & W4 & RUWE & Parallax & Distance & $A_{G}'$ & M\textsubscript{G} & G\textsubscript{BP}-G\textsubscript{RP} & Flag\\
(mag) & (mag) & (mag) & (mag) &  & [mas] & [pc] & (mag) & (mag) & (mag) &\\ 
\hline
   ... & ... & ... & ... & ... & ... & ... & ... & ... & ... & ...\\ 
   10.82 & 10.56 & 9.76 & 9.15 & 0.96 & $0.321\pm0.023$ & $2850^{+200}_{-180}$ & 2.31 & -1.78 & -0.25 & -\\ 
   
   \addlinespace[0.07cm]
   10.23 & 9.99 & 9.35 & 9.00 & 0.89 & $0.356\pm0.035$ & $2600^{+270}_{-230}$ & 1.37 & -0.78 & 0.08 & W3W4\\ 
   
   \addlinespace[0.07cm]
   10.14 & 9.83 & 8.92 & 7.70 & 1.11 & $0.286\pm0.033$ & $3130^{+360}_{-300}$ & 1.36 & -1.85 & 0.09 & -\\
   
   \addlinespace[0.07cm]
   10.63 & 10.33 & 9.40 & 8.21 & 1.30 & $0.108\pm0.044$ & $6000^{+1700}_{-1200}$ & - & - & - & -\\ 
   
   \addlinespace[0.07cm]
   9.17 & 8.91 & 8.33 & 7.50 & 1.03 & $0.256\pm0.028$ & $3470^{+390}_{-320}$ & 2.11 & -2.50 & 0.25 & W3W4\\
   
   \addlinespace[0.07cm]
   10.23 & 10.03 & 9.32 & 8.71 & 0.82 & $0.388\pm0.078$ & $2400^{+590}_{-400}$ & - & - & - & W3W4\\  
   
   \addlinespace[0.07cm]
   10.25 & 10.00 & 9.31 & 8.19 & 0.99 & $0.347\pm0.034$ & $2660^{+270}_{-230}$ & 2.47 & -2.27 & -0.07 & -\\  
   
   \addlinespace[0.07cm]
   10.68 & 9.75 & 10.42 & 8.46 & 0.95 & $1.137\pm0.032$ & $858^{+24}_{-23}$ & 1.52 & 1.14 & 0.24 & W3W4\\ 
   
   \addlinespace[0.07cm]
   10.15 & 9.53 & 9.55 & 8.86 & 1.06 & $0.864\pm0.027$ & $1121^{+35}_{-33}$ & 0.61 & 1.39 & 0.18 & W3W4\\ 
   
   \addlinespace[0.07cm]
   11.06 & 10.84 & 9.91 & 8.63 & 0.93 & $0.340\pm0.046$ & $2710^{+410}_{-320}$ & 0.81 & -1.08 & -0.06 & X-mtch\\  
   
   \addlinespace[0.07cm]
   10.47 & 10.22 & 9.40 & 8.70 & 0.90 & $0.217\pm0.036$ & $3990^{+670}_{-510}$ & 1.48 & -1.80 & 0.35 & W3W4\\ 
   
   \addlinespace[0.07cm]
   11.20 & 10.78 & 10.53 & 8.66 & 0.97 & $0.446\pm0.056$ & $2110^{+290}_{-230}$ & 0.91 & -0.16 & 0.11 & -\\    
   ... & ... & ... & ... & ... & ... & ... & ... & ... & ... & ...\\   
   
\hline      
\end{tabular}
\tablefoot{This table is available in electronic form in its entirety with uncertainties of the magnitudes, quality flags, and rest of \textit{Gaia} parameters together with angular distances from AllWISE and IPHAS or VPHAS+ sources to \textit{Gaia} sources. A portion is shown here for guidance regarding its form and content. The probabilities are expressed to the precision of their uncertainties. Distances from \citet{2018AJ....156...58B}. $A_{G}'$ only traces the interstellar extinction. It is used to correct $M_{G}$ and $G_{BP}-G_{RP}$. $A_{G}'$, $M_{G}$, and $G_{BP}-G_{RP}$ are only presented for sources with RUWE<1.4 and $\varpi/\sigma(\varpi)\geq5$. The probabilities of the three categories sum up to 1. The different catalogue warning flags are discussed throughout the text and summarised in Appendix \ref{AppendixD}. We note that some of the sources presented here have a `X-mtch' flag (57 out of 693), and hence they are likely false candidates because of an incorrect cross-match with IPHAS or VPHAS+.}
\end{table*}

\end{appendix}
\end{document}